\documentclass{article}
\usepackage{tikz}

\usepackage[utf8]{inputenc}
\usepackage{textcomp}
\usepackage{chetnew}
\usepackage{tikz}
\usepackage{amssymb,amsfonts,mathrsfs,dsfont,yfonts,bbm}\usetikzlibrary{calc}
\usepackage{xcolor}
\usepackage{braket}

\renewcommand{\tilde}{\widetilde}

\renewcommand{\hat}{\widehat}
\newcommand{\cJ}{\mathcal{J}}

\newcommand{\cW}{\mathcal{W}}
\newcommand{\cX}{\mathcal{X}}
\newcommand{\cI}{\mathcal{I}}

\usepackage{amsmath}	

\begin{document}

\preprint{YITP-17-119}
\title{On the chiral algebra of Argyres-Douglas theories\\[3mm] and S-duality}

\author{Jaewang Choi$^{\clubsuit, 1}$ and Takahiro Nishinaka$^{\heartsuit, 2}$}

\affiliation{\smallskip$^1$ Yukawa Institute for Theoretical Physics (YITP)\\
Kyoto University, Kyoto 606-8502, Japan\\
 \smallskip$^2$ Department of Physical Sciences, College of Science and Engineering \\ Ritsumeikan University, Shiga 525-8577, Japan\emails{$^{\clubsuit}$jchoi@yukawa.kyoto-u.ac.jp, $^{\heartsuit}$nishinak@fc.ritsumei.ac.jp}}

\abstract{We study the two-dimensional chiral algebra associated with the simplest Argyres-Douglas type theory with an exactly marginal coupling, i.e.,~the $(A_3,A_3)$ theory. Near a cusp in the space of the exactly marginal deformations (i.e.,~the conformal manifold), the theory is well-described by the $SU(2)$ gauge theory coupled to isolated Argyres-Douglas theories and a fundamental hypermultiplet. In this sense, the $(A_3,A_3)$ theory is an Argyres-Douglas version of the $\mathcal{N}=2$\, $SU(2)$ conformal QCD. By studying its Higgs branch and Schur index, we identify the minimal possible set of chiral algebra generators for the $(A_3,A_3)$ theory, and show that there is a unique set of closed OPEs among these generators. The resulting OPEs are consistent with the Schur index, Higgs branch chiral ring relations, and the BRST cohomology conjecture. We then show that the automorphism group of the chiral algebra we constructed contains a discrete group $G$ with an $S_3$ subgroup and a homomorphism $G\to S_4 \times {\bf Z}_2$. This result is consistent with the S-duality of the $(A_3,A_3)$ theory.}

\date{}

\maketitle
\setcounter{tocdepth}{2}
\toc

\section{Introduction}

It has recently been shown that every four-dimensional $\mathcal{N}=2$ superconformal field theory (SCFT) contains a set of BPS operators whose operator product expansions (OPEs) are characterized by a two-dimensional chiral algebra \cite{Beem:2013sza}. These BPS operators are annihilated by two Poincar\'e and two conformal supercharges, and called ``Schur operators.''
They include the highest weight component of the $SU(2)_R$ current and the Higgs branch operators, but their complete spectrum is generally highly non-trivial.  The associated chiral algebra, however, determines the complete spectrum of the Schur operators, and moreover characterizes their OPEs. Indeed, some general properties of 4d $\mathcal{N}=2$ SCFTs have been uncovered with the help of the associated chiral algebras \cite{Beem:2014zpa, Liendo:2015ofa, Lemos:2015orc, Lemos:2016xke, Beem:2016wfs}.

The 2d chiral algebra is especially a powerful tool to study the OPEs of strongly coupled SCFTs. Among other SCFTs, Arygres-Douglas type SCFTs are of particular importance. These theories are a series of $\mathcal{N}=2$ SCFTs with Coulomb branch operators of fractional scaling dimensions, whose first examples were found as IR SCFTs at singular points on the Coulomb branch of asymptotically free gauge theories \cite{Argyres:1995jj, Argyres:1995xn, Eguchi:1996vu}. Their generalizations are then constructed by wrapping M5-branes on a Riemann surface \cite{Gaiotto:2009hg, Bonelli:2011aa, Xie:2012hs} as well as by considering type II string theory on a Calabi-Yau singularity \cite{Shapere:1999xr, Cecotti:2010fi}.  While their behavior at generic points on the Coulomb branch is well-studied,\footnote{For example, the BPS index of Argyres-Douglas theories at generic points on the Coulomb branch is studied in \cite{Gaiotto:2008cd, Gaiotto:2009hg, Alim:2011ae, Alim:2011kw, Gaiotto:2012rg}.} it is very recent that their physics at the conformal point began to be understood.\footnote{The conformal anomalies and the flavor central charges were calculated long ago by the pioneering works \cite{Aharony:2007dj, Shapere:2008zf}.} In particular, guided by the Schur limit of the superconformal indices of the Argyres-Douglas theories \cite{Buican:2015ina, Cordova:2015nma, Buican:2015tda},\footnote{See \cite{Buican:2015hsa} for how the Schur indices of Argryes-Douglas theories capture the spectrum of the Coulomb branch operators in the theories despite the fact that the Coulomb branch operators do not contribute to the Schur index. See also \cite{Buican:2015tda, Song:2015wta} for other limits of the superconformal indices of the Argyres-Douglas theories.} the associated chiral algebras of a class of Argyres-Douglas theories were identified \cite{Beem:2013sza, Buican:2015ina, Cordova:2015nma, Buican:2016arp, Xie:2016evu, Creutzig:2017qyf, Song:2017oew, Buican:2017fiq}. These results are strong enough to determine the complete spectrum of the Schur operators in this class of Argyres-Douglas theories.

One common feature to the above class of Argyres-Douglas theories is, however, that they do not have an exactly marginal coupling, except for the one studied in \cite{Buican:2016arp}. In this sense, these theories are isolated in the space of 4d $\mathcal{N}=2$ SCFTs. On the other hand, there are many Argyres-Douglas type theories with an exactly marginal coupling, whose associated chiral algebras have not been identified (with an exception mentioned above).\footnote{Here, by Argyres-Douglas type theories, we mean 4d $\mathcal{N}=2$ SCFTs with Coulomb branch operators of fractional dimensions. On the other hand, the presence of an $\mathcal{N}=2$ preserving exactly marginal coupling is equivalent to that of a Coulomb branch operator of dimension two.} Since such SCFTs are in continuous families parameterized by the value of the coupling, many interesting phenomena with no counterpart in isolated SCFTs occur in them. In particular, the space of the values of exactly marginal couplings (i.e.,~the conformal manifold) often has an action of a discrete group so that theories at two different points connected by the group action are dual to each other. This duality is usually referred to as the ``S-duality.'' When the conformal manifold has weak and strong coupling points, the S-duality often exchanges them.\footnote{See \cite{Gomis:2015yaa, Donagi:2017vwh , Tachikawa:2017aux} for general discussions on the topology of the conformal manifold of 4d $\mathcal{N}=2$ SCFTs. A 4d $\mathcal{N}=1$ SCFT with a compact conformal manifold (without cusps) was found in \cite{Buican:2014sfa}.} In this sense, the S-duality is a strong/weak duality, and therefore highly non-trivial in general. 
While the S-duality of theories without fractional-dimensional Coulomb branch operators were well-studied \cite{Seiberg:1994aj, Argyres:2007cn, Gaiotto:2009we},
that of the Argyres-Douglas type SCFTs are still not very clear despite many non-trivial works \cite{Buican:2014hfa, DelZotto:2015rca, Xie:2016uqq, Xie:2017vaf, Xie:2017aqx}. In particular, the action of the S-duality group on the associated chiral algebras is an interesting open problem.

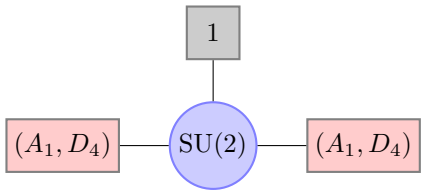
\begin{figure} 
\begin{center}
\vskip .5cm
 \begin{tikzpicture}[gauge/.style={circle, draw=blue!50, fill=blue!20, thick, inner sep=0pt, minimum size=7mm},hyper/.style={rectangle, draw=black!50, fill=black!20, thick, inner sep=0pt, minimum size=7mm},box/.style={rectangle, draw=black!50, fill=red!20, thick, inner sep=0pt, minimum size=7mm},auto]
  \node[box] (0) at (0,0) {\;$(A_1,D_4)$\;};
  \node[gauge] (1) at (2,0) {$\;\text{SU}(2)\;$} edge[-] (0);
  \node[box] (2) at (4,0) {\;$(A_1,D_4)$\;} edge[-] (1);
  \node[hyper] (3) at (2,1.5) {$1$} edge[-] (1);
\end{tikzpicture}
\caption{The quiver diagram describing the $(A_3, A_3)$ theory near a cusp in the conformal manifold. The top box with $1$ in it stands for a fundamental hypermultiplet of $SU(2)$, and the left and right boxes stand for two copies of the $(A_1,D_4)$ theory. The middle circle stands for an $SU(2)$ vector multiplet which gauges the diagonal $SU(2)$ flavor symmetry of the three sectors.}
\label{fig:quiver}
\end{center}
\end{figure}

In this paper, we study the chiral algebra associated with the simplest Argyres-Douglas type theory with an exactly marginal coupling, i.e.,~the $(A_3, A_3)$ theory \cite{Cecotti:2010fi, Xie:2012hs}.\footnote{This ``simplicity'' is based on the dimension of the Coulomb branch of the theory. In this sense, the $(A_3, A_3)$ theory is simpler than the theory studied in \cite{Buican:2016arp}.} While the $(A_3,A_3)$ theory has a Calabi-Yau singularity construction \cite{Cecotti:2010fi} and a class $\mathcal{S}$ construction \cite{Xie:2012hs}, the most convenient construction for us below is given by the quiver gauge theory described in Fig.~\ref{fig:quiver} \cite{Buican:2014hfa}; it is obtained by gauging the diagonal $SU(2)$ flavor symmetry of a fundamental hypermultiplet and two copies of an isolated Argyres-Douglas theory called $(A_1,D_4)$ theory.\footnote{The $(A_1, D_4)$ theory is sometimes referred to as $H_2$ Arygres-Douglas theory or $(A_2,A_2)$ theory. Its Seiberg-Witten (SW) curve is given by $x^2 = z^2 + b z + m_1 + \frac{u}{z} + \frac{m_2^2}{z^2}$ with the SW 1-form $\lambda = xdz$. Here $b$ and $u$ are a relevant coupling and the vacuum expectation value of a Coulomb branch operator of dimension $\frac{3}{2}$, while $m_1$ and $m_2$ are mass parameters associated with the $SU(3)$ flavor symmetry of the theory. See \cite{Argyres:1995xn, Eguchi:1996vu, Cecotti:2010fi, Bonelli:2011aa, Xie:2012hs} for more detail.} Since the beta function of this gauging vanishes, the whole theory is an $\mathcal{N}=2$ SCFT with an exactly marginal gauge coupling. In particular, the S-duality of the theory exchanges the weak and strong gauge coupling limits. According to its Seiberg-Witten curve \cite{Cecotti:2010fi, Xie:2012hs}, the theory has two Coulomb branch operators of dimension $\frac{3}{2}$ and one Coulomb branch operator of dimension $2$. The flavor symmetry of the theory is $U(1)^3$, and the conformal anomalies are given by $a= \frac{15}{8}$ and $c=2$ \cite{Xie:2012hs, Buican:2014hfa}.\footnote{We use the normalization with which a free hypermultiplet has $a=\frac{1}{24}$ and $c=\frac{1}{12}$.} Moreover, it has recently been found in \cite{Buican:2017uka, Buican:2014hfa} that the Schur index of the $(A_3,A_3)$ theory is invariant under an action of $S_4 \times {\bf Z}_2$, an $S_3$ subgroup of which is associated with the S-duality.

To identify the chiral algebra associated with the $(A_3, A_3)$ theory, we first study the Higgs branch chiral ring of the theory. 
According to \cite{Beem:2013sza}, the generators of the Higgs branch chiral ring map to generators of the associated chiral algebra.
In addition to them, the highest weight component of the $SU(2)_R$ current maps to the 2d stress tensor, which is another independent generator of the chiral algebra. While there could, in principle, be more generators of the chiral algebra, the minimal conjecture is that these are the complete set of generators of the chiral algebra. With this conjecture, we will bootstrap the OPEs among the generators by solving the constraint of the OPE associativity, or equivalently, Jacobi identities. We find that there exists a unique set of OPEs consistent with the Jacobi identities. We will then show that our result is consistent with the Schur index, Higgs branch operator relations, and the BRST cohomology conjecture made in \cite{Beem:2013sza}.

 Since the Schur index of the theory  is identical to the character of the associated chiral algebra \cite{Beem:2013sza},  the $S_4\times {\bf Z}_2$ symmetry of the Schur index is expected to have a counterpart acting on the chiral algebra. Indeed, we find that the automorphism group of the chiral algebra we constructed contains a discrete group $G$ with a group homomorphism $G\to S_4\times {\bf Z}_2$. The kernel of this homomorphism is generated by flipping the signs of operators of half-integer holomorphic dimensions, to which the Schur index/character is not sensitive.
Moreover, we show that $G$ contains an $S_3$ subgroup, which is naturally interpreted as the one associated with the S-duality. This identification is perfectly consistent with the S-duality action on the Schur index.

The organization of this paper is the following. In section \ref{sec:A3A3}, we briefly review known features of the $(A_3,A_3)$ theory, and then study its Higgs branch chiral ring. In section \ref{sec:chiral_algebra}, after giving a very quick review of \cite{Beem:2013sza}, we bootstrap the chiral algebra associated with the $(A_3, A_3)$ theory, with a minimal conjecture on the set of generators, and then give various consistency checks of our result. In sub-section \ref{subsec:S4}, we study the automorphism group of the chiral algebra we constructed. We give a conclusion and discuss future directions in section \ref{sec:Discussions}.

\section{$(A_3,A_3)$ theory}
\label{sec:A3A3}

In this section, we briefly review the S-duality and the Schur index of the $(A_3, A_3)$ Argyres-Douglas theory, and then study its Higgs branch chiral ring. The Schur index and the Higgs branch chiral ring will be important for the identification of the associated chiral algebra in the next section.

\subsection{Quiver description and S-duality}

As shown in \cite{Buican:2014hfa} and further studied in \cite{DelZotto:2015rca, Xie:2016uqq, Xie:2017vaf}, the conformal manifold of the $(A_3, A_3)$ theory has a weak coupling point at which the theory is well-described by the quiver diagram shown in Fig.~\ref{fig:quiver}. Note that each $(A_1,D_4)$ theory has $SU(3)$ flavor symmetry, an $SU(2)$ subgroup of which is now gauged. Therefore, there is a residual $U(1)$ flavor symmetry acting on each $(A_1,D_4)$ sector. Similarly, there is a $U(1)$ flavor symmetry acting on the fundamental hypermultiplet. Therefore, in total, the flavor symmetry of the $(A_3, A_3)$ theory is $U(1)^3$.
Since the beta function for the $SU(2)$ gauging vanishes, the gauge coupling $\tau = \frac{\theta}{\pi} + \frac{8\pi i}{g^2}$ is an exactly marginal coupling of the $(A_3,A_3)$ theory.

It was shown in \cite{Buican:2014hfa} that the Seiberg-Witten curve of the $(A_3,A_3)$ theory is invariant under two duality transformations $T$ and $S$, which act on the gauge coupling $\tau$ as\footnote{These transformations are denoted by $\tilde T$ and $\tilde S$ in \cite{Buican:2014hfa}.}
\begin{align}
  T:\quad \tau \to \tau + 1~,\qquad  S:\quad\tau \to -\frac{1}{\tau}~.
\end{align}
Therefore the duality group acts on the space of $\tau$ as $PSL(2,{\bf Z})$. Since it relates the weak coupling limit $\tau = i\infty$ with various strong coupling limits, the duality is referred to as the S-duality. In addition to changing the value of $\tau$, the S-duality also permutes operators of the same scaling dimensions. In particular, it permutes the flavor $U(1)^3$ currents through the symmetric group $S_3$. The permutation of the flavor currents is uniquely determined by the change of the coupling $\tau$ through the group homomorphism 
\begin{align}
PSL(2,{\bf Z}) \to S_3~.
\label{eq:hom0}
\end{align}
This is the $(A_3,A_3)$ counterpart of the $SO(8)$ triality associated with the S-duality of the $\mathcal{N}=2$ $SU(2)$ gauge theory with four flavors studied in \cite{Seiberg:1994aj}.\footnote{See also \cite{Cecotti:2015hca} for more on the S-duality action.}

\subsection{Schur index}
\label{subsec:index}

Since it will be useful in identifying the chiral algebra in Sec.~\ref{sec:chiral_algebra}, we next review the Schur index of the $(A_3,A_3)$ theory. The Schur index is generally defined by
\begin{align}
 \mathcal{I}(q;{\bf x}) \equiv \text{Tr}_{\mathcal{H}} (-1)^F q^{E-R}\prod_{i=1}^{\text{rank} G_F}(x_i)^{f_i}~,
\label{eq:Schur0}
\end{align}
where $\mathcal{H}$ is the Hilbert space of local operators, $G_F$ is the flavor symmetry of the theory, and $E, R$ and $f_i$ are the scaling dimension, $SU(2)_R$ charge and the flavor charges of the operator. In the case of $(A_3, A_3)$ theory, we have $G_F=U(1)^3$.

The quiver description reviewed above implies that the Schur index of the $(A_3,A_3)$ theory is evaluated as follows \cite{Buican:2015ina} :
\begin{align}
 \mathcal{I}_{(A_3, A_3)} &= \oint_{|w|=1} \frac{dw}{2\pi i w}\Delta(w)\;\mathcal{I}_{\text{vector}}(w)\;\cI_{\text{hyper}}(w,a)\;\mathcal{I}_{(A_1,D_4)}(w,b)\; \mathcal{I}_{(A_1,D_4)}(w,c)~,
\label{eq:A3A3-index1}
\end{align}
where $a,\,b$ and $c$ are the flavor fugacities for the $U(1)^3$ flavor symmetry.\footnote{To be more precise, $a$ is the fugacity for the $U(1)$ flavor symmetry acting on the fundamental hypermultiplet, and $b$ and $c$ are fugacities for the $U(1)$ flavor symmetries acting on the two $(A_1, D_4)$ theories.} 
The factor $\Delta(w)\equiv \frac{1}{2}(1-w^2)(1-w^{-2})$ comes from the Haar measure of $\mathfrak{su}(2)$, and
\begin{align}
 \mathcal{I}_{\text{vector}}(w)= P.E.\left[-\frac{2q}{1-q}\chi^{\mathfrak{su}(2)}_{\bf 3} (w)\right]~,\qquad \mathcal{I}_{\text{hyper}}(w,a) =  P.E.\left[\frac{\sqrt{q}}{1-q}(a+a^{-1})\chi^{\mathfrak{su}(2)}_{\bf 2}(w)\right]~,
\end{align}
are the index contributions from the $\mathfrak{su}(2)$ vector multiplet and the fundamental hypermultiplet, respectively. Here, $P.E.[g(q;x_1\cdots x_k)] \equiv \exp\left(\sum_{n=1}^\infty \frac{1}{n}g(q^n;(x_1)^n,\cdots, (x_{k})^n)\right)$ for any function, $g$, of fugacities, and $\chi_{\bf n}^{\mathfrak{su}(2)}(w)\equiv (w^{n}-w^{-n})/(w-w^{-1})$ is the character of $n$-dimensional irreducible representation of $\mathfrak{su}(2)$. The other factor, $\mathcal{I}_{(A_1,D_4)}(w,a)$, is the Schur index of the $(A_1,D_4)$ theory, which was first computed in \cite{Buican:2015ina, Cordova:2015nma} inspired by \cite{Beem:2013sza} and turns out to be equivalent to the vacuum character of $\hat{\mathfrak{su}(3)}_{-\frac{3}{2}}$. Here, we write the simplest expression for the index which was conjectured in \cite{Xie:2016evu} and proven in \cite{kac2017remark} (See also \cite{Creutzig:2017qyf, Buican:2017fiq}):
\begin{align}
\mathcal{I}_{(A_1,D_4)}(w,a) = P.E.\left[ \frac{q}{1-q^2}\chi_{\bf 8}^{\mathfrak{su}(3)}(w,a)\right]~,
\end{align}
where our convention for the $\mathfrak{su}(3)$ character is such that the character of the adjoint representation is written as $\chi^{\mathfrak{su}(3)}_{\bf 8}(w,a) = 2 + w^2+ \frac{1}{w^2} + \left(a+\frac{1}{a}\right)\left(w + \frac{1}{w}\right)$. In terms of the relabeled flavor fugacities, $x = a,\, y = \sqrt{b c}$ and $z =\sqrt{b/c}$, the Schur index of the $(A_3,A_3)$ theory is expanded as
\begin{align}
\mathcal{I}_{(A_3, A_3)}(q;x,y,z) &= 1 + 3q + \left(xyz + \frac{xy}{z} + \frac{yz}{x} + \frac{zx}{y} + \frac{x}{yz} + \frac{y}{zx} + \frac{z}{xy} + \frac{1}{xyz}\right)q^{\frac{3}{2}} 
\nonumber\\
&\qquad + \left(10 + x^2 + \frac{1}{x^2} + y^2 + \frac{1}{y^2} + z^2 + \frac{1}{z^2}\right)q^2
\nonumber\\
& \qquad + 4\left(xyz + \frac{xy}{z} + \frac{yz}{x} + \frac{zx}{y} + \frac{x}{yz} + \frac{y}{zx} + \frac{z}{xy} + \frac{1}{xyz}  \right)q^{\frac{5}{2}} + \mathcal{O}(q^3)~.
\label{eq:Schur}
\end{align}

Note that the above Schur index is invariant under the action of $S_3$ generated by
\begin{align}
 &\sigma_1: \quad x\longleftrightarrow y~,\qquad z: \text{ fixed}~,
\label{eq:S4-1}
\\
 &\sigma_2:\quad y\longleftrightarrow z~,\qquad x: \text{ fixed}~.
\label{eq:S4-2}
\end{align}
Since this $S_3$ is identical to the $S_3$ in the homomorphism \eqref{eq:hom0}, this symmetry of the index is interpreted as the S-duality invariance of the Schur index of the $(A_3,A_3)$ theory \cite{Buican:2015ina}.
In other words, the S-duality group acts on the Schur index through this $S_3$.\footnote{Note here that, since the Schur index is independent of the gauge coupling, the action of the full $PSL(2,{\bf Z})$ is not visible in it.} In addition to this $S_3$, it was shown in \cite{Buican:2017uka} that the Schur index is also invariant under
\begin{align}
 &\sigma_3: \quad x\longleftrightarrow y^{-1}~,\quad z:\text{ fixed}~,
\label{eq:S4-3}
\\
 &\zeta: \quad x\longrightarrow x^{-1}~,\quad y\longrightarrow y^{-1}~,\quad z\longrightarrow z^{-1}~.
\label{eq:S4-4}
\end{align}
Here, $\zeta$ corresponds to the charge conjugation of the theory.\footnote{To be more precise, $\zeta$ corresponds to the $CP$ transformation, namely the combination of the charge conjugation and the parity transformation. Here, since our 4d theory is a strongly coupled CFT without a conventional Lagrangian description, the two transformations, $C$ and $CP$, are purely defined as linear maps from the Hilbert space of operators to itself. In our case, $\zeta$ exchanges the chiral supercharge $\mathcal{Q}^I{}_\pm$ and the anti-chiral one $\tilde{\mathcal{Q}}_{I\dot\pm}$ since the two supercharges have opposite $U(1)_r$ and $SU(2)_R$ charges (See, for example, appendix A of \cite{Beem:2013sza}). For this to be consistent with the supersymmetry algebra $\{\mathcal{Q}^I{}_\alpha,\, \tilde{\mathcal{Q}}_{J\dot\alpha}\} = \delta^I_J \mathcal{P}_{\alpha\dot\alpha}$, the $\zeta$ must also involve the parity transformation. 
Therefore, the $\zeta$ corresponds to $CP$ instead of $C$. This point, however, does not play any crucial role in this paper.} On the other hand, $\sigma_3$ is such that $\sigma_1\sigma_3: x\to x^{-1},\, y\to y^{-1}$, which is the charge conjugation of the two $(A_1,D_4)$ theories in the quiver in Fig.~\ref{fig:quiver}. Hence, the Schur index of the $(A_3, A_3)$ theory is invariant under the action of the group generated by $\sigma_1,\sigma_2,\sigma_3$ and $\zeta$, which turns out to be $S_4 \times {\bf Z}_2$. 

According to the discussion in \cite{Beem:2013sza}, the above Schur index is equivalent to the character of the associated chiral algebra. This will be important when we identify the chiral algebra for the $(A_3,A_3)$ theory in Section \ref{sec:chiral_algebra}.

\subsection{Higgs branch from 3d reduction}

Having given a quick review of the known features of the $(A_3,A_3)$ theory, we here study the Higgs branch chiral ring of the theory.
When one space direction is compactified, the theory reduces to a 3d $\mathcal{N}=4$ theory. While the Coulomb branch is modified in this reduction, the Higgs branch is expected to be invariant. Therefore, the Higgs branch chiral ring of the $(A_3,A_3)$ theory can be read off from its 3d reduction.

\begin{figure}
\begin{center}
\vskip .5cm
 \begin{tikzpicture}[gauge/.style={circle, draw=blue!50, fill=blue!20, thick, inner sep=0pt, minimum size=7mm},hyper/.style={rectangle, draw=black!50, fill=black!20, thick, inner sep=0pt, minimum size=7mm},box/.style={rectangle, draw=black!50, fill=red!20, thick, inner sep=0pt, minimum size=7mm},auto]
  \node[gauge] (0) at (0,-.8) {\;$U(1)_a$\;};
  \node[gauge] (1) at (2,0) {\;$U(1)_d$\;} edge[-]  (0);
  \node[gauge] (2) at (4,-.8) {\;$U(1)_b$\;} edge[-] node[below]{$q_{ab},\,q_{ba}$} (0) edge[-] (1);
  \node[gauge] (3) at (2,1.8) {\;$U(1)_c$\;} edge[-] (0) edge[-] (1) edge[-] node{$q_{bc},\,q_{cb}$}(2);
\end{tikzpicture}
\caption{The quiver diagram for the 3d reduction of the $(A_3,A_3)$ theory. Each circle stands for a $U(1)$ vector multiplet, and each edge stands for a hypermultiplet charged under the $U(1)$ gauge symmetries at its ends. Note that this quiver is self-mirror, namely its 3d mirror pair is itself. Therefore, its Coulomb and Higgs branches are isomorphic to each other.}
\label{fig:quiver2}
\end{center}
\end{figure}
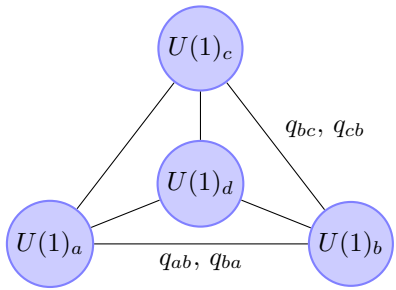

The 3d reduction of the $(A_3,A_3)$ theory is known to be the IR limit of the gauge theory described by the quiver diagram in figure \ref{fig:quiver2} \cite{Xie:2012hs}.\footnote{To be more precise, \cite{Xie:2012hs} shows that the mirror of the 3d reduction of the $(A_3,A_3)$ theory is described by theory associated with the quiver diagram in figure \ref{fig:quiver2}. Since the quiver gauge theory is self-mirror, this mirror transformation just exchanges the Coulomb and Higgs branches. The fact that this 3d quiver gauge theory is self-mirror can be seen from the general rule of abelian mirror symmetry found in \cite{deBoer:1996ck} (See Sec.~4 in particular).}
Note that we here remove the decoupled diagonal $U(1)$ of the $U(1)_a\times U(1)_b\times U(1)_c\times U(1)_d$ vector multiplets.
Let us denote by $q_{ab}$ and $q_{ba}$ the two scalar fields in the
hypermultiplet charged under $U(1)_a\times U(1)_b$. We say that, under
$U(1)_a\times U(1)_b$, $q_{ab}$ has charge $(+1,-1)$ while $q_{ba}$ has
$(-1,+1)$.\footnote{In the 3d $\mathcal{N}=4$ language, $q_{ab}$ and $q_{ba}$ are usually denoted by $q_{ab}$ and $\tilde{q}_{ab}$, respectively. We here use $q_{ab}$ and $q_{ba}$ for later convenience.} The superpotential of the 3d theory is given by
\begin{align}
 W_\text{superpot} &= (\Phi_{a}-\Phi_b)q_{ab}q_{ba} + (\Phi_b-\Phi_c)q_{bc}q_{cb} + (\Phi_a-\Phi_c)q_{ac}q_{ca}
\nonumber\\
&\qquad + (\Phi_{a}-\Phi_d)q_{ad}q_{da} + (\Phi_d-\Phi_b)q_{db}q_{bd} + (\Phi_c-\Phi_d)q_{cd}q_{dc}~.
\label{eq:superpot}
\end{align}
Note that the coefficients can be set to any non-vanishing value by rescaling the hypermultiplet scalars. Now, the Higgs branch chiral ring is the ring of gauge invariant operators composed of the hypermultiplet scalars subject to the F-term conditions.

\begin{table}
\begin{center}
 \begin{tabular}{c|c|c|c|c|c|c}
  & $q_{ab}$ & $q_{bc}$ & $q_{ac}$ & $q_{ad}$ & $q_{db}$ & $q_{cd}$ \\
\hline
$U(1)_{1}$ & & $1/2$ & $-1/2$ & $1/2$ & $1/2$ & \\
$U(1)_{2}$ & $1/2$ & & $-1/2$ & & $-1/2$ & $-1/2$ \\
$U(1)_{3}$ & $-1/2$ & $-1/2$ & & $1/2$ & & $-1/2$ \\
 \end{tabular}
\caption{The flavor charges of the hypermultiplet scalars in the 3d reduction.}
\label{table:charges}
\end{center} 
\end{table}
The theory has a flavor $U(1)^3=U(1)_{1}\times U(1)_{2}\times U(1)_{3}$ symmetry acting on the hypermultiplet scalars; the $U(1)^3$ shifts the phases of $q_{ij}$ and $q_{ji}$ to the opposite directions by the same amount. We take the three flavor charges 
as in Table~\ref{table:charges}~.
There are Higgs branch operators in the same supermultiplet as the flavor currents. They are called the ``flavor moment maps,'' and  
written as
       	 \begin{align}
       	  \cJ^1 &= -\frac{1}{2}\left(q_{bc}q_{cb} -q_{ac}q_{ca}+q_{ad}q_{da}+ q_{db}q_{bd}\right)~,
       	 \\
       	\cJ^2 &= -\frac{1}{2}\left(q_{ab}q_{ba}-q_{ac}q_{ca}-q_{db}q_{bd}-q_{cd}q_{dc} \right)~,
       \\
       	  \cJ^3 &= -\frac{1}{2}\left(-q_{ab}q_{ba}-q_{bc}q_{cb}+ q_{ad}q_{da}-q_{cd}q_{dc} \right)~.
       	 \end{align}
 Since the flavor symmetry is abelian, these moment maps are neutral under the flavor symmetry.

The Higgs branch chiral ring is generated by the above flavor moment maps and the following gauge invariant operators:
\begin{align}
 \cW_{+++} &= \frac{q_{ad}q_{dc}q_{ca}}{\sqrt{3}}~,\qquad \cW_{---} = \frac{q_{ac}q_{cd}q_{da}}{\sqrt{3}}~,\qquad \cW_{+-+} = \frac{q_{ad}q_{db}q_{ba}}{\sqrt{3}}~,\qquad \cW_{-+-} = \frac{q_{ab}q_{bd}q_{da}}{\sqrt{3}}~,
\\
\cW_{-++} &= \frac{q_{cb}q_{bd}q_{dc}}{\sqrt{3}}~, \qquad \cW_{+--} = \frac{q_{bc}q_{cd}q_{db}}{\sqrt{3}}~,\qquad \cW_{++-} = \frac{q_{ab}q_{bc}q_{ca}}{\sqrt{3}}~,\qquad \cW_{--+} = \frac{q_{ac}q_{cb}q_{ba}}{\sqrt{3}}~,
\end{align}
and
\begin{align}
 \cX^1_+ &= -\frac{\sqrt{2}}{3}q_{ad}q_{db}q_{bc}q_{ca}~,\qquad \cX^1_- = \frac{\sqrt{2}}{3}q_{ac}q_{cb}q_{bd}q_{da}~,\qquad \cX^2_+ = \frac{\sqrt{2}}{3}q_{ab}q_{bd}q_{dc}q_{ca}~,
\\
\cX^2_- &= -\frac{\sqrt{2}}{3}q_{ac}q_{cd}q_{db}q_{ba}~,\qquad \cX^3_+ = -\frac{\sqrt{2}}{3}q_{ad}q_{dc}q_{cb}q_{ba}~,\qquad \cX^3_- = \frac{\sqrt{2}}{3}q_{ab}q_{bc}q_{cd}q_{da}~.
\end{align}
Here, $\cW_{q_1q_2q_3}$ has charge $(q_1,q_2,q_3)$ under flavor $U(1)_{1}\times U(1)_{2}\times U(1)_{3}$ while $\cX^i_q$ has charge $2q$ for $U(1)_{i}$ and is neutral under $U(1)_{j}$ for $j\neq i$. Note that $\mathcal{W}$s and $\mathcal{X}$s correspond respectively to ``triangles'' and ``rhombi'' in the quiver diagram. The normalizations of the operators are chosen so that they are easily compared with null operators of the associated chiral algebra in the next section.

The above generators  satisfy the following chiral ring relations:
 \begin{align}
 2\sqrt{2}\cW_{\pm jk}\cW_{\pm mn} \pm \left(\epsilon_{jm}\epsilon_{kn}\cJ^2\cX^1_{\pm} +\cJ^3\cX^1_{\pm}\right)&=0~,
\label{eq:Higgs1}\\
 2\sqrt{2}\cW_{k\pm j}\cW_{n\pm m} \pm \left(\epsilon_{jm}\epsilon_{kn}\cJ^3\cX^2_{\pm} +\cJ^1\cX^2_{\pm} \right)&=0~,
\\
2\sqrt{2}\cW_{jk\pm}\cW_{mn\pm} \pm \left(\epsilon_{jm}\epsilon_{kn}\cJ^1\cX^3_{\pm} +\cJ^2\cX^3_{\pm} \right)&=0~,
\\
3\epsilon^{ik}\epsilon^{jl}\cW_{+ij}\cW_{-kl} + \frac{1}{2}\cJ^1\Big((\cJ^2)^2+(\cJ^3)^2\Big)&=0~,
\\
3\epsilon^{ik}\epsilon^{jl}\cW_{j+i}\cW_{l-k} + \frac{1}{2}\cJ^2\Big((\cJ^3)^2+(\cJ^1)^2\Big)&=0~,
\\
3\epsilon^{ik}\epsilon^{jl}\cW_{ij+}\cW_{kl-} + \frac{1}{2}\cJ^3\Big((\cJ^1)^2+(\cJ^2)^2\Big)&=0~,
\label{eq:Higgs2}
\end{align}
which follow from the F-term conditions associated with the superpotential \eqref{eq:superpot}. Here we used the anti-symmetric tensor such that $\epsilon^{+-} = 1$.
 In the next section, we will construct the chiral algebra for the $(A_3,A_3)$ theory which is consistent with these Higgs branch operator relations.

\section{Chiral algebra conjecture}
\label{sec:chiral_algebra}

In this section, we will construct the chiral algebra associated, in the sense of \cite{Beem:2013sza}, with the $(A_3, A_3)$ theory. To that end, we first review some results of \cite{Beem:2013sza} in Sec.~\ref{subsec:review}. We then focus on the $(A_3,A_3)$ theory, and identify the minimal possible set of chiral algebra generators in Sec.~\ref{subsec:generators}. In Sec.~\ref{subsec:bootstrap}, we bootstrap the OPEs among them by assuming the absence of extra generators. We finally give various consistency checks of our result in Sec.~\ref{subsec:Schur} -- \ref{subsec:BRST}. We follow the convention of \cite{Beem:2013sza} in this and later sections.

\subsection{Review of the chiral algebra construction}
\label{subsec:review}

Let us briefly review the chiral algebra construction of \cite{Beem:2013sza}. In this sub-section, we consider a general 4d $\mathcal{N}=2$ SCFT.
The central objects in the analysis are BPS local operators called ``Schur operators,'' which are defined by local operators annihilated by the four supercharges $\mathcal{Q}^1{}_-,\,\mathcal{S}_1{}^-,\,\tilde{\mathcal{Q}}_{2\dot-}$ and $\tilde{\mathcal{S}}^{2\dot-}$. They satisfy
\begin{align}
 E = j_1 + j_2 + 2R~,
\label{eq:Schur2}
\end{align}
where $E$ is the scaling dimension, $(j_1,j_2)$ is the $SO(4)$ spin, and $R$ is the $SU(2)_R$ charge of the operators.

The 2d chiral algebra is obtained by putting the Schur operators on ${\bf R}^2$ in the four-dimensional spacetime.\footnote{We here work in the Euclidean four-dimensional spacetime.} We then consider the ``twisted translation'' on ${\bf R}^2$ as
\begin{align}
 \mathcal{O}^\text{twisted}(z,\bar{z}) = e^{zL_{-1} + \bar{z}(\overline{L}_{-1} + R^-)}\; \mathcal{O}^\text{Schur}(0)\; e^{-zL_{-1} - \bar{z}(\overline{L}_{-1} + R^-)}~,
\end{align}
where $\mathcal{O}^\text{Schur}(0)$ is a Schur operator located at the origin of ${\bf R}^2$, $L_{-1}$ and $\overline{L}_{-1}$ are the holomorphic and anti-holomorphic translations on ${\bf R}^2$, and $R^-$ is the lowering operator of $\mathfrak{su}(2)_R$. An important feature of this twisted translation is that, $\mathcal{O}^\text{twisted}(z,\bar z)$ is closed with respect to a nilpotent linear combination of supercharges, $\mathcal{Q}\equiv \mathcal{Q}^1{}_- + \tilde{\mathcal{S}}^{2\dot-}$. Moreover, $\overline{L}_{-1} + R^-$ turns out to be $\mathcal{Q}$-exact. Therefore, the 2d OPEs of $\mathcal{Q}$-cohomology classes, $\mathcal{O}^\text{2d}(z)\equiv [\mathcal{O}^\text{twisted}(z,\bar z)]_{\mathcal{Q}}$, are meromorphic in $z$. Thus, we have a general map
\begin{align}
 \mathcal{O}^\text{Schur}(x) \;\; \longrightarrow  \;\;  \mathcal{O}^\text{2d} (z)~.
\label{eq:map}
\end{align}
The OPEs of $\mathcal{O}^\text{2d}(z)$ form a chiral algebra, or vertex operator algebra, on ${\bf R}^2$. Since the structure of the chiral algebra is completely determined by the four-dimensional OPEs of the Schur operators, identifying the chiral algebra for a given $\mathcal{N}=2$ SCFT will reveal much information of the original four-dimensional OPEs.

Let us briefly review typical examples of Schur operator. In terms of the classification of irreducible $\mathcal{N}=2$ superconformal multiplets in \cite{Dolan:2002zh}, the only multiplets containing a Schur operator are $\hat{\mathcal{B}}_R,\,\overline{\mathcal{D}}_{R(j_1,0)},\,\mathcal{D}_{R(0,j_2)}$ and $\hat{\mathcal{C}}_{R(j_1,j_2)}$ multiplets. In particular, the highest weight component of the $SU(2)_R$ current, $(\cJ_{SU(2)_R})^{ij}_{\alpha\dot\alpha}$, in the stress tensor multiplet, $\hat{\mathcal{C}}_{0(0,0)}$, is a Schur operator present in any 4d $\mathcal{N}=2$ SCFT.\footnote{Here, $i,j = 1,2$ are the symmetric $SU(2)_R$ indices for the adjoint representation of $\mathfrak{su}(2)_R$, while $\alpha=\pm$ and $\dot\alpha=\dot\pm$ are spinor indices for the vector representation of $\mathfrak{so}(4)$.} Indeed, it has $E=3,\, j_1=j_2=\frac{1}{2}$ and $R=1$, and satisfies the Schur condition \eqref{eq:Schur2}. If the theory has a flavor symmetry, the flavor current is in a $\hat{\mathcal{B}}_1$ multiplet. While the flavor current itself is not Schur, the bottom component of the $\hat{\mathcal{B}}_1$ multiplet is a Schur operator called ``flavor moment map.'' 
More generally, the bottom component of $\hat{B}_R$ multiplets are Schur operators called ``Higgs branch operators,'' 
whose vacuum expectation values parameterize the Higgs branch moduli space of vacua.\footnote{Higgs branch operators are local operators annihilated by $\mathcal{Q}^1{}_\alpha,\, \tilde{\mathcal{Q}}_{2\dot\alpha}$ and their conjugates. They saturate four unitarity bounds so that $j_1=j_2=0$ and $E=2R$.} It is generally shown in \cite{Beem:2013sza} that, by the map \eqref{eq:map}, the $(\cJ_{SU(2)_R})^{11}_{+\dot+}$ is mapped to the Virasoro stress tensor with central charge
\begin{align}
c_{2d}=-12c_{4d}~,
\end{align}
where $c_{4d}$ is the four-dimensional ``$c$ central charge.''\footnote{Here, the normalization of $c_{4d}$ is such that a free hypermultiplet contributes $c_{4d} = \frac{1}{12}$.}
Similarly, a flavor moment map is mapped to an affine current with level
\begin{align}
 k_{2d} = -\frac{k_{4d}}{2}~,
\end{align}
where $k_{4d}$ is the flavor central charge of the corresponding flavor current in four dimensions.\footnote{Our normalization of $k_{4d}$ is such that a fundamental hypermultiplet of the flavor symmetry contributes $k_{4d} = 2$.} More generally, Higgs branch operators are mapped to Virasoro primaries. 

Let us finally review the relation to the Schur index \eqref{eq:Schur0}. Since $E-R$ in \eqref{eq:Schur0} commutes with the four supercharges annihilating Schur operators, the index gets contributions only from the Schur operators. Therefore, the same quantity is calculated in terms of the chiral algebra. Indeed, it agrees with the character of the vacuum representation of the chiral algebra
\begin{align}
\text{Tr}(-1)^Fq^{L_0}\prod_{i=1}^{\text{rank} G_F} x_i{}^{J^i_0}~,
\end{align}
where the trace is taken over the Hilbert space of chiral algebra operators, and $L_0$ and $J^i_0$ are respectively the zero modes of the stress tensor and the affine current associated with the $i$-th Cartan generator of $G_F$.

\subsection{Generators}
\label{subsec:generators}

Let us now focus on the $(A_3, A_3)$ theory. According to our discussion in Sec.~\ref{sec:A3A3}, the $(A_3,A_3)$ theory has {\it at least} the following Schur operators. 
\begin{itemize}
 \item The highest weight component of $SU(2)_R$ current, $(\mathcal{J}_{SU(2)_R})^{11}_{+\dot+}$. The four-dimensional $c$ central charge is $c_{4d}=2$.
 \item $U(1)^3$ flavor moment maps $\cJ^a$. As explained in appendix \ref{app:flavor}, the corresponding flavor $U(1)^3$ current has the flavor central charge $k_{4d}=8$.
 \item Baryonic Higgs branch operators $\cW_{q_1q_2q_3}$ for $q_1,q_2,q_3=\pm 1$ and $\cX^i_{q}$ for $i=1,2,3$ and $q=\pm 1$.\footnote{In this paper, we call flavor-charged Higgs branch operators which are not flavor moment maps ``baryons.''} The $\cW_{q_1q_2q_3}$ has charge $(q_1,q_2,q_3)$ under $U(1)^3$, and $\cX^i_q$ has charge $2q$ under $U(1)_i$. They are subject to the Higgs branch chiral ring relations in \eqref{eq:Higgs1} -- \eqref{eq:Higgs2}.
\end{itemize}
In particular, the last two bullets correspond to the generators of the Higgs branch chiral ring.
Here, we use the same notation for the four-dimensional Higgs branch operators as for their three-dimensional reductions. 
According to the dictionary reviewed above, these Schur operators map to the following operators of the chiral algebra.
\begin{itemize}
\item The stress tensor, $T$, with the central charge $c_{2d}$ such that
\begin{align}
 T(z)T(0) \sim \frac{c_{2d}}{2z^4} + \frac{2T}{z^2} + \frac{\partial T}{z}~.
\label{eq:OPE-T}
\end{align}
The central charge is given by $c_{2d}=-12c_{4d} = -24$.
 \item An affine $U(1)^3$ current, $J^i$, at level $k_{2d}$ such that
\begin{align}
 J^i(z)J^j(0) \sim \frac{k_{2d}\delta^{ij}}{2z^2}~,
\label{JJOPE}
\end{align}
for $i,j=1,\cdots,3$. The level is given by $k_{2d} = -\frac{1}{2}k_{4d} = -4$.
\item A primary operator $W_{q_1q_2q_3}$ for $q_i=\pm 1$. It is of dimension $\frac{3}{2}$ and has $U(1)^3$ charge $(q_1,q_2,q_3)$. Therefore
\begin{align}
 T(z)W_{q_1q_2q_3}(0) &\sim \frac{\frac{3}{2}W_{q_1q_2q_3}}{z^2} + \frac{\partial W_{q_1q_2q_3}}{z}~,
\\
J^i(z)W_{q_1q_2q_3}(0) &\sim \frac{q_i W_{q_1q_2q_3}}{z}~,
\label{JWOPE}
\end{align}
\item A primary operator $X^i_q$ of dimension $2$ for $i=1,2,3$ and $q=\pm1$. It has charge $2q$ under $U(1)_i$. Therefore
\begin{align}
 T(z) X^i_q(0) &\sim \frac{2X^i_q}{z^2} + \frac{\partial X^i_q}{z}~,
\\[2mm]
J^j(z) X^i_q(0) &\sim \frac{2q\;\delta^{ij}\,X^i_q}{z}\qquad \text{not summed over }i~.
\label{XJOPE}
\end{align}
\end{itemize}

Note that these operators are independent generators of the chiral algebra. In other words, there is no smaller set of operators whose OPEs generate all of $T,\,J^i,\,W_{q_1q_2q_3}$ and $X_q^i$. Indeed, it was shown in \cite{Beem:2013sza} that the generators of the Higgs branch chiral ring are mapped to independent generators of the associated chiral algebra. In addition, the stress tensor is also an independent generator since the Sugawara stress tensor of affine $U(1)^3$ currents cannot reproduce $c_{2d}=-24$.\footnote{The Virasoro central charge $c_{2d}$ of the Sugawara stress tensor of an affine $U(1)^n$ current is given by $c_{2d}=n$ regardless of the level of the current. In particular, it cannot be negative.}

The chiral algebra could, in principle, contain more generators. The minimal possible assumption is, however, that those associated with the $SU(2)_R$ current and the Higgs branch generators are the full set of generators of the chiral algebra. This minimal assumption is known to be applicable to various $\mathcal{N}\geq 2$ SCFTs such as a series of 4d rank-one $\mathcal{N}=2$ SCFTs \cite{Beem:2013sza}, a certain class of isolated Argyres-Douglas theories \cite{Beem:2013sza, Xie:2016evu, Creutzig:2017qyf, Buican:2017fiq}, and several 4d $\mathcal{N}>2$ SCFTs \cite{Beem:2013sza, Nishinaka:2016hbw}.\footnote{See \cite{Lemos:2016xke} for more general discussion on the chiral algebras of 4d $\mathcal{N}>2$ SCFTs.}
We therefore {\it conjecture} that the chiral algebra for the $(A_3, A_3)$ theory is generated by $J^i(z),\,T(z),\,W_{q_1q_2q_3}(z)$ and $X^i_q(z)$. Below, we show that this conjecture uniquely fixes the OPEs among the generators (up to the normalizations). We also give various consistency checks of the resulting OPEs.

\subsection{OPEs among the baryonic generators}
\label{subsec:bootstrap}

Every chiral algebra satisfies the basic consistency condition of associativity, or equivalently, the Jacobi identities among the generators.\footnote{In terms of the OPEs, the Jacobi identity is expressed as
\begin{align}
 [\mathcal{O}_1(z_1) [\mathcal{O}_2(z_2) \mathcal{O}_3(z_3)]] - [\mathcal{O}_3(z_3) [\mathcal{O}_1(z_1)\mathcal{O}_2(z_2)]] - [\mathcal{O}_2(z_2)[\mathcal{O}_3(z_3)\mathcal{O}_1(z_1)]]=0~,
\end{align}
for $|z_2-z_3| < |z_1 -z_3|$, where $[\cdots]$ stands for the singular part of the OPE of the operators.} This Jacobi identity condition puts a strong constraint on the possible form of the OPEs of the generators. For our conjecture to make sense, there has to be at least one set of OPEs among $W_{q_1q_2q_3}$ and $X_q^i$ which are consistent with the Jacobi identities.

To see that this is indeed the case, we concretely solve this Jacobi identity constraint. Our strategy is the following. We first write down the most general ansatz for the OPEs among the generators which are consistent with the charge conservation. Then we impose the constraint that all the Jacobi identities are satisfied, which gives rise to various relations among the OPE coefficients in the ansatz. 
Interestingly, we find that there is one and only one set of OPE coefficients consistent with these relations. For this calculation, we used the Mathematica package \verb|OPEdefs| developed by \cite{Thielemans:1991uw}.

Moving the detailed description of this calculation to appendix \ref{app:Jacobi}, we only give the resulting OPEs here. Since the OPEs of the forms  $J^i(z) \mathcal{O}(0)$ and $T(z)\mathcal{O}(0)$ are fixed by the symmetry, we only have to describe $W_{q_1q_2q_3}(z)W_{p_1p_2p_3}(0),\, W_{q_1q_2q_3}(z)X_q^i(0)$ and $X_{q}^i(z)X_{p}^j(0)$. To write down a simple expression for them, let us define  $\tilde J^A,\, \tilde W^A_{\pm}$ and $\tilde X^{[AB]} = - \tilde X^{[BA]}$ for $A,B=1,\cdots,4$ by
 \begin{align}
 \tilde J ^1 \equiv J^1 -J^2 -J^3~,&\qquad \tilde J^2 \equiv J^2 - J^3-J^1~,\qquad \tilde J^3 \equiv J^3 - J^1 - J^2~,\qquad \tilde J^4 \equiv J^1 + J^2 + J^3~,
 \\[2mm]
 \tilde W^1_{\pm} \equiv& W_{\pm\mp\mp}~,\qquad \tilde W^2_{\pm} \equiv W_{\mp\pm\mp}~,\qquad \tilde W^3_{\pm} \equiv W_{\mp\mp\pm}~,\qquad \tilde W^4_{\pm} \equiv W_{\pm\pm\pm}~,
\label{eq:rewritingW}
\\[2mm]
   \tilde X^{[i4]} &\equiv X^i_{+}~,\qquad \tilde X^{[ij]} \equiv -\epsilon^{ij}{}_{k4}X^k_{-} \quad \text{for} \quad i,j,k=1,2,3~,
\label{eq:rewritingX}
\end{align}
where $\epsilon^{ABCD}$ is the anti-symmetric tensor such that $\epsilon^{1234}=1$, and raising/lowering the indices is realized by $\delta^{AB}$ and $\delta_{AB}$.
In terms of these variables, the only set of OPEs consistent with the Jacobi identities is written as
\begin{align}
 \tilde W^A_{\pm}(z) \tilde W^B_{\mp}(0) &\sim \pm \delta^{AB}\left\{\frac{1}{z^3} \mp\frac{\tilde J^A}{2z^2} + \frac{1}{z}\left(-\frac{T}{6} - \frac{1}{96}\tilde{J}_E\tilde J^E + \frac{1}{8}\left(\tilde J^A\right)^2 \mp \frac{1}{4}\left(\tilde J^A{}'\right)\right)\right\}~,
\label{eq:OPE1}
\\
\tilde W^A_{+}(z)\tilde W^B_{+}(0) &\sim 
-\frac{\tilde X^{[AB]}}{\sqrt{2}z}~,
\label{eq:WW+}
\\
\tilde W^A_{-}(z)\tilde W^B_{-}(0) &\sim \frac{\epsilon^{AB}{}_{EF} \tilde X^{[EF]}}{2\sqrt{2}z}~,
\label{eq:WW-}
\\
\tilde W^A_{+}(z) \tilde X^{[BC]}(0) &\sim \epsilon^{ABC}{}_E\frac{1}{\sqrt{2}}\left(\frac{\tilde W^E_-}{z^2} - \frac{\left(\tilde J^{A} -\frac{1}{2}\tilde J^{B}-\frac{1}{2}\tilde J^C\right)\tilde W^{E}_-}{3z} + \frac{\tilde W^{E}_-{}'}{3z}\right)~,
\\
\tilde W^A_{-}(z) \tilde X^{[BC]}(0) &\sim -\delta^{A[B}\delta^{C]}_{E} \sqrt{2}\left(\frac{\tilde W^E_+}{z^2} - \frac{\left(\tilde J^{A} +\frac{1}{2}\tilde J^{B}+\frac{1}{2}\tilde J^C\right)\tilde W^{E}_+}{3z} + \frac{\tilde W^{E}_+{}'}{3z}\right)~,
\\
\tilde X_{[AB]}(z) \tilde X^{[CD]}(0) &\sim -\epsilon_{AB}{}^{CD}\Bigg[\frac{1}{z^4} - \frac{\tilde J_A + \tilde J_B}{2z^3}  + \frac{1}{z^2}\left(-\frac{2}{9}T+ \frac{1}{8}(\tilde J_A + \tilde J_B)^2 - \frac{1}{4}(\tilde J_A +\tilde J_B)'-\frac{1}{72}\tilde J_E \tilde J^E \right)
\nonumber\\
&\quad  + \frac{1}{z}\bigg(
 \frac{1}{54}T(\tilde J_A + \tilde J_B) - \frac{7}{216}(\tilde J_A + \tilde J_B)^3 +\frac{1}{54}(\tilde J_A + \tilde J_B)\tilde J_E \tilde J^E
\nonumber\\
&\qquad\qquad   + \frac{8}{9}\{\tilde W_{A+},\tilde W_{A-}\} +\frac{8}{9}\{\tilde W_{B-},\tilde W_{B+}\}-\frac{5}{18}\{\tilde W_{E\,-},\tilde W^E_+\}  +\frac{1}{8}(\tilde J_A +\tilde J_B) (\tilde J_A +\tilde J_B){}'
\nonumber\\
&\qquad \qquad \qquad  - \frac{1}{72}\tilde J_E \tilde J^E{}'-\frac{1}{9}T'\bigg)\Bigg] +  \frac{2}{3z}\epsilon_{[A}{}^{CDF}\{\tilde W_{B]\,+},\,\tilde W_{F\,-}\}~,
\label{eq:OPEs}
\end{align}
where the indices $A,B,C,D$ are {\it not} summed while $E$ and $F$ are summed over $1,\cdots,4$.
For the detailed derivation of this result, see appendix \ref{app:Jacobi}.
In the rest of this section, we will give several consistency checks of this result.

\subsection{Consistency with the Schur index}
\label{subsec:Schur}

Let us first check that the above OPEs are consistent with the Schur index of the $(A_3,A_3)$ theory. To that end, we compute the character of the chiral algebra we constructed.
Since our chiral algebra is generated by $J^i,T,\, W_{q_1q_2q_3}$ and $X^i_q$,
if there are no null operators then the character of the chiral algebra is given by
\begin{align}
 P.E.\left[ \frac{1}{1-q}\left(3q + q^2 + q^{\frac{3}{2}}\sum_{i,j,k=\pm 1} x^{i}y^{j}z^k + q^2\sum_{i=\pm 1}(x^{2i} + y^{2i} + z^{2i})\right) \right]
\label{eq:PE}
\end{align}
where the first two terms are the contributions of $J^i$ and $T$, the third is that of $W_{q_1q_2q_3}$, and the fourth is that of $X^i_q$. However, there are indeed various null operators in the chiral algebra. Therefore, the correct character is obtained by subtracting the contributions of the null operators from \eqref{eq:PE}. Note here that, while the chiral algebra is non-unitary, every null operator has to be removed since every chiral algebra operator corresponds to a Schur operator in a {\it unitary} four-dimensional SCFT. Moreover, as explained in Appendix \ref{app:Jacobi}, removing the null operators is necessary for the Jacobi identities to be satisfied.
Below, we identify null operators of dimension less than or equal to $4$, and then show that the character obtained this way coincides with the Schur index of the $(A_3, A_3)$ theory evaluated in \eqref{eq:Schur}.

The first non-trivial null operators appear at dimension $3$. There are indeed 12 null operators charged under the flavor symmetry, and 3 neutral null operators. Their index contribution is given by
\begin{align}
 q^3\left(2x^2 + \frac{2}{x^2} + 2y^2 + \frac{2}{y^2} + 2z^2 + \frac{2}{z^2} + 3\right)~.
\label{eq:null-index1}
\end{align}
At dimension $7/2$, there are 3 null operators with charge $(\pm1,\pm1, \pm1)$. Their index contribution is
\begin{align}
3 q^{7/2}\left(xyz + \frac{1}{xyz} + \frac{xy}{z} + \frac{z}{xy} + \frac{yz}{x}+\frac{x}{yz} + \frac{zx}{y} + \frac{y}{zx}\right)~.
\label{eq:null-index2}
\end{align}
At dimension $4$, there are 18 neutral null operators. 
In addition to them, there is one null operator for all charge combinations of the form $(q_1,q_2,0),\,(0,q_2,q_3)$ and $(q_1,0,q_3)$ with $q_k = \pm 2$. 
The last set of null operators of dimension four are 9 null operators for each of charge $(\pm 2,0,0),\,(0,\pm 2,0)$ and $(0,0,\pm2)$.
In total, the index contribution of the null operators of dimension four is given by
\begin{align}
&q^4 \biggl(18+  \bigl(x^2 +\frac{1}{x^2}\bigr)\bigl(y^2 +\frac{1}{y^2}\bigr) +\bigl(y^2 +\frac{1}{y^2}\bigr)\bigl(z^2+\frac{1}{z^2}\bigr) +\bigl(z^2+\frac{1}{z^2}\bigr) \bigl(x^2 +\frac{1}{x^2}\bigr)
\nonumber\\
&\qquad + 9 \biggl(x^2 +\frac{1}{x^2}+ y^2 +\frac{1}{y^2}+z^2+\frac{1}{z^2}\biggr)  \biggr)~.
\label{eq:null-index2}
\end{align}

Subtracting \eqref{eq:null-index1} -- \eqref{eq:null-index2} from \eqref{eq:PE}, we obtain the character of the chiral algebra up to $\mathcal{O}(q^{4})$, which turns out to be in perfect agreement with the Schur index of the $(A_3,A_3)$ theory. 
This result is a highly non-trivial consistency check of our result.

\subsection{Consistency with the Higgs branch chiral ring}

The OPEs \eqref{eq:OPE1} -- \eqref{eq:OPEs} imply that there are the following null operators of dimension 3:
\begin{align}
& 2\sqrt{2}W_{\pm jk}W_{\pm mn}  \pm \left(\epsilon_{jm}\epsilon_{kn}J^2X^1_{\pm} +J^3X^1_{\pm} - \epsilon_{kn}X^1_{\pm}{}'\right)\sim 0~,
\label{eq:null1}
\\
& 2\sqrt{2}W_{k\pm j}W_{n\pm m} \pm \left(\epsilon_{jm}\epsilon_{kn}J^3X^2_{\pm} +J^1X^2_{\pm} - \epsilon_{kn}X^2_{\pm}{}'\right)\sim 0~,
\\
& 2\sqrt{2}W_{jk\pm}W_{mn\pm} \pm \left(\epsilon_{jm}\epsilon_{kn}J^1X^3_{\pm} +J^2X^3_{\pm} - \epsilon_{kn}X^3_{\pm}{}'\right)\sim 0~,
\\
&3\epsilon^{ik}\epsilon^{jl}W_{+ij}W_{-kl} +\frac{1}{2}J^1\Big((J^2)^2+(J^3)^2\Big) - TJ^1 +T'- J^AJ^A{}'+\frac{3}{2}J^1{}'{}'\sim 0~,
\\
&3\epsilon^{ik}\epsilon^{jl}W_{j+i}W_{l-k} +\frac{1}{2}J^2\Big((J^3)^2+(J^1)^2\Big) - TJ^2 +T'- J^AJ^A{}'+\frac{3}{2}J^2{}'{}'\sim 0~,
\\
&3\epsilon^{ik}\epsilon^{jl}W_{ij+}W_{kl-} +\frac{1}{2}J^3\Big((J^1)^2+(J^2)^2\Big) - TJ^3 +T'- J^AJ^A{}'+\frac{3}{2}J^3{}'{}'\sim 0~.
\label{eq:null2}
\end{align}
These null operators correspond to operator relations in four dimensions. Indeed, when the terms involving the stress tensor or derivative are eliminated, the above null operators are precisely identical to the Higgs branch relations in \eqref{eq:Higgs1} -- \eqref{eq:Higgs2}. Hence, the above null operators of the chiral algebra we constructed are perfectly consistent with the Higgs branch operator relations of the $(A_3,A_3)$ theory.\footnote{Let us stress again that every null operator in the chiral algebra has to be removed even though the chiral algebra is non-unitary. The reason is that every chiral algebra operator corresponds to a Schur operator in four dimensions, in which no local operator is null. Moreover, as explained in Appendix \ref{app:Jacobi}, removing the null operators is necessary for the Jacobi identities.}

The stress tensor and derivatives of operators have no counterparts in \eqref{eq:Higgs1} -- \eqref{eq:Higgs2}. The reason for this is that the former is not associated with a Higgs branch operator in four-dimensions, and the latter is descendants of some primary operators and therefore identified with zero in the chiral ring. Similar phenomena happen in many 4d rank-one $\mathcal{N}=2$ SCFTs \cite{Beem:2013sza} and 4d rank-one $\mathcal{N}>2$ SCFTs \cite{Beem:2013sza, Nishinaka:2016hbw}.

\subsection{Consistency with BRST cohomology}
\label{subsec:BRST}

Recall that the $(A_3,A_3)$ theory is obtained by an exactly marginal gauging of isolated theories, as described by the quiver diagram in Fig.~\ref{fig:quiver}. In such a case, the chiral algebra is conjectured, in \cite{Beem:2013sza}, to be obtained by a certain type of BRST cohomology of the direct sum of chiral algebras for the constituents of the quiver. When the gauge coupling is turned on, some of the Schur operators in the constituents are lifted to non-Schur. 
The BRST cohomology is then expected to remove all such lifted operators as those that are not BRST-closed or those that are BRST-exact. Below, we will check that our conjecture is consistent with this BRST cohomology conjecture. 

Let us illustrate how the BRST cohomology is constructed in our case. The discussion below directly follows from the general analysis in \cite{Beem:2013sza}. We first write down the direct sum of the chiral algebras of the constituents. Since the quiver diagram is composed of a fundamental hypermultiplet of $SU(2)$, two $(A_1,D_4)$ theories, and an $SU(2)$ vector multiplet, the direct sum of the corresponding chiral algebras is generated by symplectic bosons $q^k_s$, two copies of $\widehat{\mathfrak{su}(3)}_{-\frac{3}{2}}$ current, $j^P_n$, and the $b^I,\partial c^J$ ghosts in the adjoint representation of $\mathfrak{su}(2)$. Their OPEs are written as
\begin{align}
j^P_n(z) \, j^Q_m(0)  \sim  \delta_{nm}\left(\frac{-3/2 \; \delta^{PQ}}{2z^2} + \frac{i f^{PQ}{}_R \, j^R_m }{z}\right)~, \qquad 
q_s^k(z) \, q_t^\ell(0) \sim  \frac{\delta_{st} \, \epsilon^{k\ell} }{z}~, \qquad 
b^I(z)\,c^J(0) \sim \frac{\delta^{IJ}}{z}~,
\end{align}
where $n,m=1,2$ specifies in which $(A_1,D_4)$ theory the current $j^P_n$ is included, the $P,\,Q,\,R = 1,\cdots, 8$ are the adjoint indices of $\mathfrak{su}(3)$, and the $f^{PQ}{}_R$ is the structure constant of $\mathfrak{su}(3)$. The $k,\ell=1,2$ and $I,J=1,2,3$ are respectively the indices for the fundamental and adjoint representations of $\mathfrak{su}(2)$, and the $s$ and $t$ run over $1,2$. We raise and lower the $\mathfrak{su}(3)$ and $\mathfrak{su}(2)$ indices respectively by $\delta_{PQ},\, \delta^{PQ}$ and $\delta_{IJ},\,\delta^{IJ}$.

Note that each $\widehat{\mathfrak{su}(3)}$ current is decomposed into an $\widehat{\mathfrak{su}(2)}\times \widehat{\mathfrak{u}(1)}$ current and two doublets of $\mathfrak{su}(2)$, the latter of which are charged oppositely under the $\mathfrak{u}(1)$. We take the basis of the $\widehat{\mathfrak{su}(3)}$ current so that $f^{PQR} = \epsilon^{PQR}$ and $f^{PQ8} = 0$ for $P,Q,R=1,2,3$. Therefore $J^I_n$ for $I=1,2,3$ and $J^8_n$ form an $\widehat{\mathfrak{su}(2)}\times \hat{u(1)}$ sub-algebra of $\widehat{\mathfrak{su}(3)}_{-\frac{3}{2}}$.\footnote{This decomposition corresponds to the decomposition of the ${\bf 8}$ multiplet of $\mathfrak{su}(3)$ into ${\bf 3}_0 \oplus {\bf 2}_1 \oplus {\bf 2}_{-1} \oplus {\bf 1}_0$ as a representation of $\mathfrak{su}(2)\times \mathfrak{u}(1)$. In terms of $j^P_n$, ${\bf 3}$ corresponds to $j^I_n$, ${\bf 1}_0$ corresponds to $J^8_n$, and ${\bf 2}_1 \oplus {\bf 2}_{-1}$ corresponds to the linear combinations which will be written in \eqref{defdoublet}.
}

In addition to the two $\widehat{\mathfrak{su}(2)}$ currents $J^I_1$ and $J^I_2$ arising from the $(A_1,D_4)$ sectors, there are also $\widehat{\mathfrak{su}(2)}$ currents built out of $q^k_s$ and $b,c$:
\begin{align}
J_\text{hyper}^I \equiv \delta^{st} q_s^k\, (\sigma^I)_{k\ell}\,  q_t^\ell~, \quad J_\text{gh}^I \equiv - i \epsilon^{I}{}_{JK}(c^J b^K)~,
\end{align}
where $I=1,2,3$. In terms of these four $\widehat{\mathfrak{su}(2)}$ currents, the BRST current is defined by
\begin{align}
J_\text{BRST}= c_I \left( j_1^I +j_2^I + J_\text{hyper}^I +\frac{1}{2} J_\text{gh}^I \right)~.
\end{align}
Then the BRST charge $Q_{\text{BRST}} \equiv \oint dz\, J_{\text{BRST}}$ turns out to be nilpotent, which reflects the fact that the $\beta$-function of the $SU(2)$ gauge coupling vanishes in four dimensions.

With respect to $Q_\text{BRST}$, we consider the BRST cohomology of the space of operators composed of $j^I_n,\,q^k_s,\,b$ and $\partial c$. To identify the basis of the cohomology, we have used the Mathematica package \verb|OPEdefs| developed in \cite{Thielemans:1991uw}. The result is the following.
At dimension one, 
there are three
independent cohomology classes represented by
\begin{align}
J^1 = \frac{2}{\sqrt{3}} (j_1^8+j_2^8)~,\qquad
J^2 = \frac{2}{\sqrt{3}} (j_1^8-j_2^8)~, \qquad 
J^3 = i \epsilon_{k\ell}\, q_1^k q_2^\ell~.
\label{eq:2d-currents}
\end{align}
The OPEs among these operators coincide with $\eqref{JJOPE}$. At dimension $\frac{3}{2}$, there are eight independent
cohomology classes represented by
\begin{align}
W_{+++} &= \frac{i}{\sqrt{6}} \epsilon_{k\ell} 
\,v_1^k (q_1^\ell+i q_2^\ell)~,\quad
W_{++-} = \frac{i}{\sqrt{6}} \epsilon_{k\ell}  
\,v_1^k (q_1^\ell-i q_2^\ell)~,\\
W_{+-+} &= \frac{i}{\sqrt{6}} \epsilon_{k\ell} 
\,v_2^k (q_1^\ell+i q_2^\ell)~,\quad
W_{+--} = \frac{i}{\sqrt{6}} \epsilon_{k\ell} 
\,v_2^k (q_1^\ell-i q_2^\ell)~, \\
W_{-++} &= \frac{i}{\sqrt{6}} \epsilon_{k\ell} 
\,u_2^k (q_1^\ell+i q_2^\ell)~, \quad
W_{-+-} = \frac{i}{\sqrt{6}} \epsilon_{k\ell} 
\,u_2^k (q_1^\ell-i q_2^\ell)~, \\
W_{--+} &= \frac{i}{\sqrt{6}} \epsilon_{k\ell} 
\,u_1^k (q_1^\ell+i q_2^\ell)~,\quad
W_{---} = \frac{i}{\sqrt{6}} \epsilon_{k\ell} 
\,u_1^k (q_1^\ell-i q_2^\ell)~,
\end{align}
where 
$u_n^i$ and 
$v_n^i$ are doublets of $\mathfrak{su}(2)$ defined as
\begin{equation}
u_n \equiv
\begin{bmatrix}
j^4_n -i j^5_n \\ 
j^6_n -i j^7_n
\end{bmatrix}
, \qquad 
v_n \equiv
\begin{bmatrix}
j^6_n +i j^7_n \\
-j^4_n -i j^5_n
\end{bmatrix}~.
\label{defdoublet}
\end{equation}
With the above normalizations of $W_{q_1q_2q_3}$, the OPEs among $W_{q_1q_2q_3}$ and $J^i$ for $i=1,2,3$ coincide with \eqref{eq:OPE1} and \eqref{JWOPE}.

At dimension two,
there are eighteen
independent cohomology classes,
two of which are null operators. 
The two null operators actually result from a null operator in 
 $\widehat{\mathfrak{su}(3)}_{-\frac{3}{2}}$ \cite{Beem:2013sza, Arakawa:2015, Beem:2017ooy}.
In terms of the rank-three invariant tensor of $\mathfrak{su}(3)$, $d^{PQR}$, these two null operators are expressed as
\begin{align}
d_{8 PQ} \, j_1^P j_1^Q~,\qquad d_{8 PQ} \, j_2^P j_2^Q~.
\end{align}
Therefore, there are sixteen independent cohomology classes up to null operators. Among them, those charged under the $\widehat{\mathfrak{u}(1)}^3$ symmetry are represented by
\begin{align}
X^1_{+} &= \frac{\sqrt{2}}{3} \epsilon_{k\ell} 
v_1^k 
v_2^\ell~,\quad
X^1_{-} = \frac{\sqrt{2}}{3} \epsilon_{k\ell} 
u_1^k 
u_2^\ell~,  \quad
X^2_{+} = \frac{\sqrt{2}}{3} \epsilon_{k\ell} 
v_1^k 
u_2^\ell~,  \quad
X^2_{-} = \frac{\sqrt{2}}{3}  \epsilon_{k\ell} 
u_1^k 
v_2^\ell~,\\
X^3_{+} &= \frac{\sqrt{2}}{3} 
\left(-\frac{1}{4}  j_1^I (q_1+i q_2)^i (\sigma_I)_{ij} (q_1+iq_2)^j   +\frac{1}{4}  j_2^I (q_1+i q_2)^i (\sigma_I)_{ij} (q_1+iq_2)^j  \right)~, \\
X^3_{-} &= \frac{\sqrt{2}}{3} 
\left(-\frac{1}{4}  j_1^I (q_1-i q_2)^i (\sigma_I)_{ij} (q_1-iq_2)^j   +\frac{1}{4}  j_2^I (q_1-i q_2)^i (\sigma_A)_{ij} (q_1-iq_2)^j  \right)~.
\end{align}
With this normalization, the OPEs among $X^a_q$ and $J^a$ for $a=1,2,3$ coincide with \eqref{XJOPE} and \eqref{eq:OPEs}.
On the other hand, those neutral under the $\widehat{\mathfrak{u}(1)}$ symmetry are represented by $J^iJ^j$, $J^i{}'$ for $i,j=1,2,3$ and
\begin{align}
T &= \frac{1}{3} \delta_{PQ}(j^P_1 j_{1}^Q+j^P_2 j_2^Q) + \frac{1}{2} \delta^{st} \epsilon_{k\ell} (q_s^k q_t^\ell) -(b^I \partial c_I)~.
\label{eq:stress}
\end{align}
Note that \eqref{eq:stress} is the sum of the stress tensors in the chiral algebras for the constituents. We see that \eqref{eq:stress} satisfies \eqref{eq:OPE-T} with $c_{2d}=-24$.

Hence, we have checked that the BRST cohomology conjecture precisely reproduces the operator spectrum of the chiral algebra we constructed, up to dimension two. In particular, we found that there are no fermionic BRST cohomology class up to dimension two.\footnote{Even though the Schur index has no term with negative sign, in principle, there could be cancellations between bosonic and fermionic contributions. However, as far as the BRST cohomology conjecture is correct, there are no such cancellations at least up to $\mathcal{O}(q^2)$, and therefore there are only bosonic Schur operators up to dimension two.} 
Moreover, we have checked that the OPE coefficients fixed by the Jacobi identities are precisely those obtained by the BRST cohomology analysis. This is a highly non-trivial consistency check of our conjecture on the chiral algebra.

\section{Automorphisms and S-duality}
\label{subsec:S4}

Having given various consistency checks of our conjecture, we here study the automorphisms of the chiral algebra. In particular, we study how the S-duality of the $(A_3, A_3)$ theory acts on the chiral algebra through its automorphism group.

It turns out that the OPEs of the chiral algebra we constructed in the previous section are invariant under the following transformations:\footnote{According to \eqref{eq:rewritingW}, the $\sigma_i$ are equivalent to 
\begin{align}
 &\hat{\sigma}_1:\quad J^1 \longleftrightarrow J^2~,\quad W_{\pm\mp\mp} \longleftrightarrow W_{\mp\pm\mp}~,\quad W_{\mp\mp\pm} \longrightarrow -W_{\mp\mp\pm}~,\quad X^1_{\pm} \longleftrightarrow X^2_\pm~,\quad X^3_\pm \longrightarrow -X^3_\pm~,
\\
 &\hat{\sigma}_2:\quad J^2 \longleftrightarrow J^3~,\quad W_{\mp\pm\mp} \longleftrightarrow W_{\mp\mp\pm}~,\quad W_{\pm\mp\mp} \longrightarrow -W_{\pm\mp\mp}~,\quad X^2_{\pm} \longleftrightarrow X^3_\pm~,\quad X^1_\pm \longrightarrow -X^1_\pm~,
\\
&\hat{\sigma}_3 : \quad J^1 \longleftrightarrow -J^2~,\quad W_{\pm\pm\pm}\longleftrightarrow W_{\mp\mp\pm}~,\quad W_{\pm\mp\mp} \longrightarrow -W_{\pm\mp\mp}~,\quad X^1_\pm \longleftrightarrow -X^2_\mp~,\quad X^3_\pm \longrightarrow -X^3_\pm~.
\end{align}
} 
\begin{align}
\hat{\sigma}_1&:\quad  \tilde J^1 \longleftrightarrow \tilde J^2~,\quad \tilde W^1_{\pm} \longleftrightarrow \tilde W^2_{\pm}~,\quad \tilde W^3_\pm \longrightarrow  -\tilde W^3_\pm~,
\nonumber\\
&\qquad \tilde X^{[13]} \longleftrightarrow -\tilde X^{[23]}~, \quad \tilde X^{[14]} \longleftrightarrow \tilde X^{[24]}~,\quad \tilde X^{[12]}\longrightarrow \tilde X^{[21]}~,\quad \tilde X^{[34]} \longrightarrow -\tilde X^{[34]}~,
\label{eq:sigma1}
\\[2mm]
\hat{\sigma}_2&:\quad  \tilde J^2 \longleftrightarrow \tilde J^3~,\quad \tilde W^2_{\pm} \longleftrightarrow \tilde W^3_{\pm}~,\quad \tilde W^4_\pm \longrightarrow  -\tilde W^4_\pm~,
\nonumber\\
&\qquad \tilde X^{[24]} \longleftrightarrow -\tilde X^{[34]}~, \quad \tilde X^{[12]} \longleftrightarrow \tilde X^{[13]}~,\quad \tilde X^{[23]}\longrightarrow  \tilde X^{[32]}~,\quad \tilde X^{[14]} \longrightarrow -\tilde X^{[14]}~,
\\[2mm]
\hat{\sigma}_3&:\quad  \tilde J^3 \longleftrightarrow \tilde J^4~,\quad \tilde W^3_{\pm} \longleftrightarrow \tilde W^4_{\pm}~,\quad \tilde W^1_\pm \longrightarrow  -\tilde W^1_\pm~,
\nonumber\\
&\qquad \tilde X^{[13]} \longleftrightarrow -\tilde X^{[14]}~, \quad \tilde X^{[23]} \longleftrightarrow \tilde X^{[24]}~,\quad \tilde X^{[34]}\longrightarrow  \tilde X^{[43]}~,\quad \tilde X^{[12]} \longrightarrow -\tilde X^{[12]}~.
\label{eq:sigma3}
\end{align}
Here, $\hat{\sigma}_i$ exchanges $i$ and $i+1$ in the anti-symmetric indices of the generators, and then flip the sign of operators with $i+2$ in their anti-symmetric indices.\footnote{For $i=3$, $\hat{\sigma}_3$ flips the sign of operators with $1$ in their indices instead.} Note that this latter sign flip is necessary for the chiral algebra to be invariant.

 In addition to the above transformations, the chiral algebra is also invariant under 
\begin{align}
 \hat{\zeta}: \quad \tilde J^A \longrightarrow -\tilde J^A~,\qquad \tilde W^A_\pm \longrightarrow i\tilde W_{\mp}^A~,\qquad \tilde X^{[AB]} \longleftrightarrow \frac{1}{2}\epsilon^{AB}{}_{CD}\tilde{X}^{[CD]}~.
\label{eq:zeta}
\end{align}
This $\hat{\zeta}$ is related to the charge conjugation invariance of the four-dimensional theory. While the charge conjugate $\mathcal{O}^\dagger$ of a Schur operator $\mathcal{O}$ is not Schur unless $\mathcal{O}={\bf 1}$, there exists a Schur operator $\tilde{\mathcal{O}}$ which sits in the same $SU(2)_R$ multiplet as $\mathcal{O}^\dagger$.\footnote{Every Schur operator $\mathcal{O}\neq {\bf 1}$ is the highest weight component of a non-trivial $SU(2)_R$ multiplet. Then $\mathcal{O}^\dagger$, which is the lowest weight component of the conjugate $SU(2)_R$ multiplet, is not Schur. In other words, the Schur sector is not invariant under the charge conjugation. However, the highest weight component $\tilde{\mathcal{O}}$ of this conjugate $SU(2)_R$ multiplet is a Schur operator.} The above $\hat{\zeta}$ is the two-dimensional version of the map $\mathcal{O}\to \tilde{\mathcal{O}}$. In other words, $\hat{\zeta}$ arises from the combination of the four-dimensional charge conjugation and a multiple action of the $SU(2)_R$ raising operator.

Thus, the chiral algebra we have constructed is invariant under the action of the group generated by $\hat{\sigma}_1, \hat{\sigma}_2, \hat{\sigma}_3$ and $\hat{\zeta}$, which we denote by $G$. The generators of the group $G$ satisfy
\begin{align}
 (\hat{\sigma}_i)^2 = {\bf 1}~,\qquad (\hat{\sigma}_1\hat{\sigma}_2)^3 = (\hat{\sigma}_2\hat{\sigma}_3)^3=(\hat{\sigma}_1\hat{\sigma}_3)^2 = \hat{\zeta}^2~,\qquad \hat{\sigma}_i\hat{\zeta} = \hat{\zeta}\hat{\sigma}_i~,\qquad \hat{\zeta}^4 = {\bf 1}.
\end{align}
The group $G$ has a normal ${\bf Z}_2$ subgroup generated by
\begin{align}
\hat{\zeta}^2: \quad W^A_\pm \longrightarrow -W^A_\pm~,
\label{eq:zeta-square}
\end{align}
which is the map that flips the signs of operators of half-integer holomorphic dimensions. Such a map is an automorphism of any chiral algebra associated with a 4d $\mathcal{N}=2$ SCFT, since the OPEs of such a chiral algebra are always single-valued \cite{Beem:2013sza}.\footnote{For chiral algebras associated with 4d $\mathcal{N}=2$ SCFTs, the holomorphic dimension of an operator is an integer or half-integer. Moreover the 2d OPEs are always single-valued. Therefore, in the OPE $\mathcal{O}_1(z)\mathcal{O}_2(0) \sim \lambda_{123}\mathcal{O}_3(0)/z^{h_1+h_2-h_3}$, if the holomorphic dimension $h_1$ of $\mathcal{O}_1$ is half-integer, one and only one of $h_2$ and $h_3$ is half-integer. If $h_3$ is half-integer, one and only one of $h_1$ and $h_2$ is half-integer. Therefore, flipping the signs of operators of half-integer holomorphic dimensions is always an automorphism of the chiral algebra. In four dimensions, $\hat{\zeta}^2$ corresponds to flipping the signs of operators of half-integer $SU(2)_R$ charges, which is also a symmetry of every $\mathcal{N}=2$ SCFT due to the nature of the tensor products of $\mathfrak{su}(2)$ representations.}
The quotient of $G$ by this ${\bf Z}_2$ is isomorphic to $S_4\times {\bf Z}_2$, namely
\begin{align}
 G/{\bf Z}_2 \simeq S_4\times {\bf Z}_2~,
\label{eq:quotient}
\end{align}
where the $S_4$ is generated by $[\hat{\sigma}_1],[\hat{\sigma}_2]$ and $[\hat{\sigma}_3]$ while ${\bf Z}_2$ is generated by $[\hat{\zeta}]$.\footnote{Here, as usual, $[g]\in G/{\bf Z}_2$ is the equivalence class represented by $g\in G$.} Note that the above quotient implies a group homomorphism
\begin{align}
 G \to S_4 \times {\bf Z}_2~.
\label{eq:hom2}
\end{align}
 Since $[\hat{\sigma}_i]$ and $[\hat{\zeta}]$ precisely induce the transformations \eqref{eq:S4-1} -- \eqref{eq:S4-4} on the Schur index, this homomorphism explains the $S_4\times {\bf Z}_2$ symmetry of the index. Therefore, we naturally identify $[\hat{\sigma}_i]$ with $\sigma_i$, and $[\hat{\zeta}]$ with $\zeta$. The kernel of \eqref{eq:hom2} is generated by $\hat{\zeta}^2$, to which the Schur index is not sensitive.\footnote{Indeed, the index just captures the spectrum of operators and is not sensitive to the signs of operators.}

Another important feature of the group $G$ is that it contains an $S_3$ subgroup. Indeed, $\hat{\sigma}_1$ and $\tilde{\sigma}_2 \equiv \hat{\sigma}_2 \hat{\zeta}^2$ generate an $S_3$ subgroup of $G$.  Then there is a homomorphism
\begin{align}
 PSL(2,{\bf Z}) \to S_3\hookrightarrow G~.
\label{eq:hom3}
\end{align}
We interpret this result to mean that the S-duality of the $(A_3,A_3)$ theory acts on its chiral algebra through the homomorphism \eqref{eq:hom3}. This interpretation is consistent with the action of the S-duality on the Schur index. Indeed, as reviewed in Sec.~\ref{subsec:index}, the S-duality acts on the Schur index through $S_3$ generated by $\sigma_1$ and $\sigma_2$. Since $[\sigma_1]$ and $[\tilde{\sigma}_2] = [\hat{\sigma}_2]$ are identified with $\sigma_1$ and $\sigma_2$, this $S_3$ is precisely the image of the $S_3$ generated by $\hat{\sigma}_1$ and $\tilde{\sigma}_2$ under the homomorphism \eqref{eq:hom2}. Therefore, the action of the S-duality on the chiral algebra and the Schur index is characterized by the chain of homomorphisms
\begin{align}
 PSL(2,{\bf Z}) \to G \to S_4\times {\bf Z}_2~.
\end{align}

\section{Discussions}
\label{sec:Discussions}

In this paper, we have studied the chiral algebra of the simplest Argyres-Douglas theory with an exactly marginal coupling, i.e., the $(A_3, A_3)$ theory. We have conjectured that the complete set of chiral algebra generators is the minimal one, i.e., they are associated with the generators of the Higgs branch chiral ring and the highest weight component of the $SU(2)_R$ current. Then we have shown that there exists a unique set of OPEs of the generators which are consistent with the Jacobi identities. Our result is consistent with the Schur index, Higgs branch operator relations, and the BRST cohomology conjecture made in \cite{Beem:2013sza}. After these consistency checks, we have studied the automorphism group of the chiral algebra we constructed, and shown that the algebra is invariant under the action of $G$ generated by \eqref{eq:sigma1} -- \eqref{eq:sigma3} as well as \eqref{eq:zeta}. This group $G$ is equipped with a group homomorphism $G\to S_4\times {\bf Z}_2$, which relates it to the symmetry of the Schur index. We have also shown that $G$ contains an $S_3$ subgroup, through which $PSL(2,{\bf Z})$ can act on the chiral algebra.

Our work is the first non-trivial step to understanding the action of the S-duality on the chiral algebras of Argyres-Douglas type theories with exactly marginal couplings. There are many open problems to be studied, including the following:
\begin{itemize}
 \item Is there a Hamiltonian reduction of an affine current algebra which reproduces the chiral algebra we have constructed in section \ref{sec:chiral_algebra}? Such a reduction was found in \cite{Creutzig:2017qyf} for $(A_1, A_{2n-1})$ and $(A_1, D_{2n})$ Argyres-Douglas theories.

 \item In the BRST cohomology analysis given in Sec.~\ref{subsec:BRST}, is there a systematic way of identifying the cohomology classes of any given holomorphic dimension? Finding such a systematic way will be useful in proving our chiral algebra conjecture for the $(A_3, A_3)$ theory.

 \item A UV $\mathcal{N}=1$ ``Lagrangian'' for the $(A_3,A_3)$ theory was conjectured in \cite{Agarwal:2017roi, Benvenuti:2017bpg} by generalizing earlier works \cite{Maruyoshi:2016tqk, Maruyoshi:2016aim, Agarwal:2016pjo, Benvenuti:2017kud}. It would be interesting to study if the discrete symmetry $G$ of the chiral algebra is visible in this UV ``Lagrangian.''

 \item Is the chiral algebra we constructed related to a quantum integrable model? The chiral algebras for the $(A_N, A_M)$ theories without flavor symmetry have been shown to be related to the quantum integrable models obtained by the ODE/IM correspondence \cite{Ito:2017ypt}. It would be interesting to study how our chiral algebra for the $(A_3, A_3)$ theory is related to this work.

 \item It would be interesting to study how the chiral algebra we constructed for the $(A_3,A_3)$ theory is related to the recent discussions on the orbifold indices of 4d $\mathcal{N}=2$ SCFTs \cite{Fredrickson:2017yka, Fluder:2017oxm, Imamura:2017wdh}. 

 \item What is the general relation between the symmetry of the chiral algebra and that of the superconformal index for general $\mathcal{N}=2$ SCFTs? Are the two symmetries identical whenever the chiral algebra contains no operator of half-integer holomorphic dimensions?

 \item What is the generalization to the $(A_N, A_N)$ theory for $N>3$? The $(A_1, A_1)$ theory is the theory of a single free hypermultiplet, whose chiral algebra is that of a symplectic boson. The chiral algebra of the $(A_2,A_2)$ theory is conjectured to be $\widehat{\mathfrak{su}(3)}_{-\frac{3}{2}}$ \cite{Beem:2013sza, Buican:2015ina, Cordova:2015nma}. The $(A_3,A_3)$ theory have been studied in this paper. It would then be interesting to study their generalization to the whole series of the $(A_N, A_N)$ theories.

\end{itemize}

In the rest of this section, we give a brief comment on the last bullet. 
According to the four-dimensional analysis in \cite{Xie:2012hs}, the $(A_N, A_N)$ theory has $c_{4d} = \frac{1}{24}N(N^2+3N-2)$, $U(1)^N$ flavor symmetry, and various baryonic Higgs branch operators. Then a natural conjecture on the chiral algebra of the $(A_N, A_N)$ theory is that it is generated by the stress tensor with $c_{2d} = -\frac{1}{2}N(N^2+3N-2)$, the affine $U(1)^N$ current, and Virasoro primaries corresponding to the baryonic Higgs branch generators.\footnote{We thank M.~Buican for having let us notice this possible conjecture and shared with us his understanding of the Higgs branch chiral rings of these theories even before we finished our analysis of the $(A_3, A_3)$ chiral algebra.} It would be interesting to study whether the constraint of the Jacobi identities fixes the OPEs among these generators as in the case of the $(A_3, A_3)$ theory.

While a detailed analysis of the $(A_N, A_N)$ chiral algebra is beyond the scope of this paper, we briefly comment on its symmetry. A crucial point is that the Schur index of the $(A_N, A_N)$ theory is invariant under the action of $S_{N+1} \times {\bf Z}_2$ \cite{Buican:2017uka}. Here $S_{N+1}$ is related to the S-duality of the theory while ${\bf Z}_2$ corresponds to the charge conjugation (together with an action of the $SU(2)_R$ raising operator). This symmetry of the Schur index suggests that the automorphism group of the chiral algebra contains a group $G$ equipped with a group homomorphism
\begin{align}
\varphi:\quad  G \to S_{N+1}\times {\bf Z}_2~.
\label{eq:hom}
\end{align}
For the $(A_3, A_3)$ theory, such a homomorphism is identified in \eqref{eq:quotient}. 
Let us denote the generators of $S_{N+1}$ by $\sigma_1,\cdots, \sigma_{N}$ and that of ${\bf Z}_2$ by $\zeta$ so that
\begin{align}
 \sigma_i^2 =(\sigma_i\sigma_{i+1})^3 = \zeta^2 = {\bf 1}~,\qquad (\sigma_i\sigma_j)^4 = {\bf 1} \quad \text{for}\quad |i-j|>1~, \qquad \sigma_i\zeta = \zeta\sigma_i~.
\end{align}
Since $\varphi$ is a homomorphism, for each generator $g$ of $S_{N+1}\times {\bf Z}_2$, there exists $\hat{g}\in G$ such that $\varphi(\hat{g}) = g$. Then, identifying the actions of $\hat{\sigma}_1,\cdots, \hat{\sigma}_N$ and $\hat{\zeta}$ on the chiral algebra generators will strongly constrain the possible OPEs of them, since the set of the OPEs has to be invariant under the action of $G$. This would be useful in constructing the chiral algebras of the $(A_N, A_N)$ theory for $N>3$.

\bigskip\bigskip\bigskip
\ack{ \bigskip
We are particularly grateful to Matthew Buican for various illuminating and inspiring discussions and communications. T.~N. also thanks him for a lot of collaborations on related topics in which T.~N. learned many techniques used in this paper. We also thank Jin-Beom Bae, Chiung Hwang, Sungjay Lee, Tomoki Nosaka, Minkyu Park, Kazuma Shimizu, Yuji Sugawara, Shigeki Sugimoto, Seiji Terashima and Kento Watanabe for useful discussions.
We thank the Yukawa Institute for Theoretical Physics for hosting the international conference, ``Strings and Fields 2017,'' at which we have useful discussions on related topics.
}

\newpage
\begin{appendices}

\section{Flavor central charges of the $(A_3,A_3)$ theory}
\label{app:flavor}
 
Here we describe how to read off the flavor $U(1)^3$ central charge $k_{4d} = 8$ of the $(A_3, A_3)$ theory. Since the theory is described at a weak coupling point by the quiver diagram in Fig~\ref{fig:quiver}, there are three mutually independent $U(1)$ currents. Let us denote by $\hat{J}^1$ the current for the $U(1)$ flavor symmetry acting on one of the two $(A_1,D_4)$ theory, by $\hat{J}^2$ that for the $U(1)$ acting on the other $(A_1,D_4)$, and by $\hat{J}^{3}$ that for the $U(1)$ acting on the fundamental hypermultiplet. Then
\begin{align}
 \hat{J}_\mu^1(x) \hat{J}_\nu^1 (0) &\sim \frac{3k_1}{4\pi^4}\frac{g_{\mu\nu}x^2 -2x_\mu x_\nu}{x^8} + \cdots ~,
\\
 \hat{J}_\mu^2(x) \hat{J}_\nu^2 (0) &\sim \frac{3k_1}{4\pi^4}\frac{g_{\mu\nu}x^2 - 2x_\mu x_\nu}{x^8}+\cdots ~,
\\
 \hat{J}_\mu^3(x) \hat{J}_\nu^3 (0) &\sim \frac{3k_2}{4\pi^4}\frac{g_{\mu\nu}x^2 - 2x_\mu x_\nu}{x^8}+\cdots~.
\end{align}
We normalize $\hat{J}^3_\mu$ so that the hypermultiplet scalars have charge $\pm 1$, which implies $k_2=8$.\footnote{Our normalization of the flavor central charge is that a hypermultiplet in the fundamental representation of the flavor symmetry contributes $k=2$.} On the other hand, $\hat{J}^1_\mu$ and $\hat{J}^2_\mu$ are normalize so that
\begin{align}
 \hat{J}^1_\mu(x) = \frac{2}{\sqrt{3}}J_\mu^{(1)8}(x)~, \qquad \hat{J}^2_\mu(x) = \frac{2}{\sqrt{3}}J_\mu^{(2)8}(x)~,
\label{eq:A1D4-flavor}
\end{align}
where $J^{(i)P}_\mu(x)$ for $P=1,\cdots,8$ are the flavor $SU(3)$ current of the $i$-th $(A_1,D_4)$ theory so that
\begin{align}
 J^{(i)P}_\mu(x) J^{(j)Q}_\nu(0) \sim \delta^{ij}\delta^{PQ}\frac{3k_{SU(3)}}{4\pi^4}\frac{g_{\mu\nu}x^2- x_\mu x_\nu}{x^8} + \delta^{ij}\frac{2}{\pi^2}f^{PQ}{}_R\frac{x_\mu x_\nu x^\rho J_\rho^{(i)R} (0)}{x^6} + \cdots~.
\end{align}
The identification \eqref{eq:A1D4-flavor} is the four-dimensional origin of the two-dimensional identification \eqref{eq:2d-currents} discussed in section \ref{subsec:BRST}.
Since the $SU(3)$ flavor central charge of the $(A_1,D_4)$ theory is given by $k_{SU(3)}=3$ \cite{Aharony:2007dj, Shapere:2008zf}, the identification \eqref{eq:A1D4-flavor} implies that $k_1 = 4$.

Let us now define
\begin{align}
 J^1_\mu(x)\equiv \hat{J}^1_\mu(x) + \hat{J}^2_\mu(x)~,\quad J^2_\mu(x) \equiv \hat{J}^1_\mu(x) - \hat{J}^2_\mu(x) ~,\quad J^3_\mu(x) \equiv \hat{J}^3_\mu(x) ~.
\end{align}
Then the OPEs among these currents are
\begin{align}
 J^i_\mu(x) J^j_\nu(0) \sim \delta^{ij}\frac{3k_{4d}}{4\pi^4}\frac{g_{\mu\nu}x^2 - 2x_\mu x_\nu}{x^8} + \cdots~,
\end{align}
with $k_{4d} = 8$, where $i,j=1,2,3$. This $U(1)^3$ currents are in the same supermultiplet as the $U(1)^3$ flavor moment maps, which are mapped to the affine $U(1)^3$ current $J^i(z)$ with level $k_{2d}=-\frac{1}{2}k_{4d} = -4$.

\section{Derivation of the 2d OPEs}
\label{app:Jacobi}

According to our conjecture, the chiral algebra for the $(A_3,A_3)$ theory is generated by the stress tensor $T(z)$ for $c_{2d} = -24$, the affine $U(1)^3$ current $J^a(z)$ with level $k_{2d}=-4$, and the baryonic generators $W_{q_1q_2q_3}(z)$ and $X^{a}_{q}(z)$. The OPEs of $T$ and $\mathcal{O}$ for $\mathcal{O} = T, J^a, W_{q_1q_2q_3},X^a_q$ as well as those of $J^a$ and $\mathcal{O}$ for $\mathcal{O} = J^b, W_{q_1q_2q_3},X^a_q$ are complete fixed by the conformal and flavor central charges. For the other OPEs, we here write down the most general expressions and then impose the constraint that all the Jacobi identities among the generators are satisfied. We will see that this constraint uniquely fixes those OPEs among the baryonic generators.

Let us first write down the most general OPEs. The most general OPE of $W_{q_1q_2q_3}(z)$ and $W_{q_1'q_2'q_3'}(0)$ is written as
\begin{align}
 W_{q_1q_2q_3}(z)W_{q_1'q_2'q_3'}(0) &\sim \epsilon_{q_1q_1'}\epsilon_{q_2q_2'}\epsilon_{q_3q_3'}\left(\frac{1}{z^3} + \frac{\mathcal{O}^{(1)}_{q_1q_2q_3}}{z^2} + \frac{\mathcal{O}^{(2)}_{q_1q_2q_3}}{z}\right) 
\nonumber\\
& + \left(\delta_{q_1q_1'}\epsilon_{q_2q_2'}\epsilon_{q_3q_3'}f^{(1)}_{q_1q_2q_3q_1'q_2'q_3'}\frac{X^1_{q_1}}{z}  + \;\;\text{permutations of }1,2,3\;\right) ~,
\label{eq:WW}
\end{align}
where we define $\epsilon_{+-} = 1$  and $\epsilon_{-+}=-1$, and normalize $W$s so that the most singular term is given by $\epsilon_{q_1q_1'}\epsilon_{q_2q_2'}\epsilon_{q_3q_3'}/z^3$. The $\mathcal{O}^{(1)}_{q_1q_2q_3}$ is a linear combination of the $U(1)^3$ currents, $J^i$, that depends on $q_1,q_2$ and $q_3$. On the other hand, the $\mathcal{O}^{(2)}$ is a linear combination of flavor neutral operators of dimension 2 which are built out of $T, J^i, W_{p_1p_2p_3}$ and their descendants. Note that $X^i_q$ does not appear in $\mathcal{O}^{(2)}_{q_1q_2q_3}$ since it is charged under the flavor symmetry. On the other hand, $f^{(1)}_{q_1q_2q_3q_1'q_2'q_3'}$ is a constant that could potentially vanish but turns out to be non-vanishing for the Jacobi identities to be satisfied.

The most general OPE of $W_{q_1q_2q_3}$ and $X^i_q$ is written as
\begin{align}
 W_{q_1q_2q_3}(z) X^i_q(0) &\sim \delta^{i1}\epsilon_{q_1q}\left(f^{(2)}_{q_1q_2q_3}\frac{W_{qq_2q_3}}{z^2} + \frac{\mathcal{O}^{(3)}_{qq_2q_3}}{z}\right)
+
\delta^{i2}\epsilon_{q_2q}\left(f^{(3)}_{q_1q_2q_3}\frac{W_{q_1qq_3}}{z^2} + \frac{\mathcal{O}^{(4)}_{q_1qq_3}}{z}\right)
\nonumber\\
&\qquad +
\delta^{i3}\epsilon_{q_3q}\left(f^{(4)}_{q_1q_2q_3}\frac{W_{q_1q_2q}}{z^2} + \frac{\mathcal{O}^{(5)}_{q_1q_2q}}{z}\right)~,
\label{eq:WX}
\end{align}
where $f^{(i)}_{p_1p_2p_3}$ for $i=2,3,4$ are functions of $p_1,p_2$ and $p_3$, and $\mathcal{O}^{(k)}_{p_1p_2p_3}$ for $k=3,4,5$ are linear combinations of $J^1W_{p_1p_2p_3},\,J^2W_{p_1p_2p_3},\,J^3W_{p_1p_2p_3}$ and $W_{p_1p_2p_3}'$ whose coefficients depend on $p_1,p_2$ and $p_3$.

Finally, the most general OPE of $X^i_{q_1}(z)$ and $X^j_{q_2}(0)$ is written as
\begin{align}
X^i_{q_1}(z) X^j_{q_2}(0) &\sim \delta^{ij}|\epsilon_{q_1q_2}|\left(\frac{1}{z^4} + \frac{\mathcal{O}^{(6)a}_{q_1}}{z^3} + \frac{\mathcal{O}^{(7)a}_{q_1}}{z^2} + \frac{\mathcal{O}^{(8)a}_{q_1}}{z}\right)
\nonumber\\
& + \epsilon^{1ij}f^{(5)}_{q_1q_2}\frac{W_{+\,q_1q_2}W_{-\,q_1q_2}}{z}  + \epsilon^{2ij}f^{(6)}_{q_1q_2}\frac{W_{q_2\,+\,q_1}W_{q_2\,-\,q_1}}{z}  + \epsilon^{3ij}f^{(7)}_{q_1q_2}\frac{W_{q_1q_2\,+}W_{q_1q_2\,-}}{z}~,
\label{eq:XX}
\end{align}
where we normalize $X^i_{q}$ so that the most singular term in the above OPE is given by $\delta^{ij}|\epsilon_{q_1q_2}|/z^4$. The $f^{(k)}_{q_1q_2}$ for $k=5,6,7$ are functions of $q_1$ and $q_2$, and $\mathcal{O}^{(k)}_{q_1}$ for $k=6,7,8$ are flavor neutral operators built out of $T,J^i$ and $W_{p_1p_2p_3}$.

The above three sets of OPEs are subject to the constraint of Jacobi identities.
Let us now describe how this constraint fixes the above undetermined functions $f^{(k)}$ and operators $\mathcal{O}^{(k)}$ in the above expressions. First of all, the Jacobi identities among $T,J^i$ and $W_{q_1q_2q_3}$ completely fix $\mathcal{O}^{(1)}_{q_1q_2q_3}$ and $\mathcal{O}^{(2)}_{q_1q_2q_3}$.

The next step is to see that the Jacobi identities for $W_{q_1q_2q_3},\, W_{q_1'q_2'q_3'}$ and $X^i_q$ demands that $f^{(1)}$ in \eqref{eq:WW} are non-vanishing. Then, rescaling $W_{q_1q_2q_3}$ as
\begin{align}
 W_{\pm\pm\pm}\to \alpha^{\pm1} W_{\pm\pm\pm}~,\quad W_{\pm\pm\mp}\to \beta^{\pm1} W_{\pm\pm\mp}~,\quad W_{\pm\mp\pm}\to\gamma^{\pm1} W_{\pm\mp\pm}~,\quad W_{\mp\pm\pm}\to \delta^{\pm1} W_{\mp\pm\pm}~,
\end{align}
we set the functions $f^{(1)}$ so that \eqref{eq:WW+} and \eqref{eq:WW-} are realized. Note that our original ansatz for the OPEs are invariant under this rescaling. In other words, we fix this rescaling degrees of freedom by fixing $f^{(1)}$ so that \eqref{eq:WW+} and \eqref{eq:WW-} are realized.

Now, the last step is to study the other Jacobi identities. One subtlety here is that such a computation involves operators of dimension three, some of which could potentially be null. Since the Jacobi identities have to be satisfied up to null operators, we have to identify all such null operators involved in (the LHS of) the Jacobi identities. Since operators of dimension smaller than three turn out to be not null, our strategy is that we first impose the constraint that all operators of dimension smaller than three involved in (the LHS of) the Jacobi identities vanish, and then identify the null operators of dimension three under this constraint. With this strategy, we find the null operators of dimension three shown in \eqref{eq:null1}. Since the OPEs of these operators with every chiral algebra generator contains operators of dimension larger than or equal to their own dimensions, these operators are all null. Note that this null property just follows from the constraints that the LHS of the Jacobi identities do not contain operators of dimension smaller than three. 

Finally, we solve the constraint that all the remaining Jacobi identities are satisfied up to the null operators of dimension three listed in \eqref{eq:null1}. This constraint is solved by tuning the values of $f^{(k)}$ and the linear combinations of operators, $\mathcal{O}^{(k)}$. Note here that such a tuning is possible if and only if the null operators of dimension three are subtracted. In other words, the Jacobi identities can be satisfied if and only if the null operators are eliminated. 

The above analysis fixes all the OPE coefficients in \eqref{eq:WW} and \eqref{eq:WX}. The OPE coefficients in \eqref{eq:XX} are not completely fixed, but all the undetermined coefficients are coupled only with null operators identified in \eqref{eq:null1}. Therefore, up to null operators, the OPEs among the generators are completely fixed by the constraint that the Jacobi identities among the generators are satisfied. The resulting OPEs are summarized in \eqref{eq:OPE1} -- \eqref{eq:OPEs}, in terms of \eqref{eq:rewritingW} and \eqref{eq:rewritingX}.

It would also be useful to stress here that, if we do not fix the value of $c_{2d}$ while fixing $k_{2d}=-4$, the constraint that the Jacobi identities are satisfied implies $c_{2d}=-24$. Therefore, with the flavor central charge $k_{2d}=-4$ fixed, the chiral algebra generated by $T,\, J^i,\, W_{q_1q_2q_3}$ and $X^i_q$ exists as a consistent W-algebra if and only if $c_{2d}=-24$.

\end{appendices}

\newpage
\bibliography{chetdocbib}

\end{document}